\documentclass{ar-1col}
\usepackage{cite}
\usepackage{amsmath}
\usepackage{url}
\usepackage[utf8]{inputenc} 
\usepackage{caption}
\usepackage{subcaption}

\jname{Xxxx. Xxx. Xxx. Xxx.}
\jvol{AA}
\jyear{YYYY}
\doi{10.1146/((please add article doi))}

\newcommand{\bea}{\begin{eqnarray}}
\newcommand{\eea}{\end{eqnarray}}
\newcommand{\gmc}{\gamma_\text{MC}}
\newcommand{\gmr}{\gamma_\text{MR}}
\newcommand{\br}{\mathbf{r}}
\newcommand{\bk}{\mathbf{k}}

\newcommand{\bv}{\mathbf{v}}

\begin{document}

\markboth{Fritz and Scaffidi}{Hydrodynamic phenomena in electronic transport}

\title{Hydrodynamic electronic transport}

\bibliographystyle{ar-style4.bst}

\author{L. Fritz$^1$ and T. Scaffidi$^{2,3}$
\affil{$^1$Institute for Theoretical Physics, Utrecht University, 3584CE Utrecht, The Netherlands; email: l.fritz@uu.nl}
\affil{$^2$ Department of Physics and Astronomy, University of California, Irvine, California 92697, USA}
\affil{$^3$ Department of Physics, University of Toronto, 60 St. George Street, Toronto, Ontario, M5S 1A7, Canada}
}

\begin{abstract}
The ``flow'' of electric currents and heat in standard metals is diffusive with electronic motion randomized by impurities. However, for ultraclean metals, electrons can flow like water with their flow being described by the equations of hydrodynamics. While theoretically postulated, this situation was highly elusive for decades. In the last decade, several experimental groups have found strong indications for this type of flow, especially in graphene-based devices. In this review, we give an overview of some of the recent key developments, both on the theoretical and experimental side.
\end{abstract}

\maketitle

\tableofcontents

\section{Introduction}\label{Sec:Introduction}

There are very few ‘universal truths’ in physics. Hydrodynamic behavior is one of them. The motion of any substance at high enough temperature follows the laws of hydrodynamics. 
Hydrodynamics in its original context describes the viscous motion of water. However, its principles apply in a much wider setting: In the physics of stars and interstellar matter, magnetohydrodynamics of plasmas, but also the dynamics of soft active matter. It can also be encountered in applied disciplines, including engineering: ocean dynamics, weather modeling, aviation, the dynamics of gas flowing through pipes, or traffic flow, to name a few examples. They even apply to the physics of the early universe: At energies high enough to melt protons and neutrons, the constituent quarks form the quark-gluon plasma. When a particle collider creates this state, it only lives a tiny fraction of a second. However, during that short spell, it moves according to the laws of fluid mechanics. 

The reason for this almost unreasonable versatility is the underlying simplicity and generality: the basis of hydrodynamics is the relaxation of conserved quantities towards local equilibrium. The conserved quantities are  fundamental: mass, momentum, and energy (and charge in charged systems). 

 For classical hydrodynamics, the set of differential equations that describes flow phenomena is composed of the continuity equations and the Navier-Stokes equation~\cite{Landaufluidmechanics}. The role of the latter in the description of fluid motion is comparable to the linear Maxwell-equations in electrodynamics. Taken together with appropriate boundary conditions, they describe all features of viscous flow. While these equations are basic, they are non-linear and their solution is highly non-trivial: Proving some properties of their solutions is one of the seven Millennium Prize problems in Mathematics~\cite{milleniumproblems}.

 A fundamental characteristic of fluids is their viscosity: water flows faster than honey due to its lower viscosity while having a similar density. Some classical fluids are so viscous, they appear solid. The viscosity of pitch is $10^{11}$ times that of water. In the quantum world, one also encounters viscous liquids in strongly interacting systems: the quark-gluon plasma is estimated to have a dynamic viscosity ~$10^{16}$ times that of water, thereby rivaling glass. The corresponding density of the system is enhanced by the same factor compared to water, meaning the ratio of dynamical viscosity to density in water and the quark-gluon plasma are actually comparable.

In the context of condensed matter physics, hydrodynamics has a successful history. It is applied in the description of strongly interacting one-dimensional systems, spin-excitations in insulators, as well as the dynamics in the vicinity of quantum critical points~\cite{subir}. As a rule of thumb, hydrodynamic behavior is most likely to be encountered and discussed in strongly correlated systems. 

Metals fall into the class of weakly correlated systems. However, metals are also the systems where we most often talk about the 'flow' of charges and electrical currents. Until quite recently, this 'flow' was fundamentally different from the flow of water: It resembled the erratic movement of balls on a tilted nail board with random position of the nails. 

In this review, we summarize recent developments theoretical and experimental progress in (semi-)metals in which this is not the case and the flow of electrons resembles the flow of water through a tube with all the associated phenomena.

\subsection{Scope}

This paper reviews the advances in the field of electronic hydrodynamics that have taken place over the last decade from the point of view of a theorist. It is geared towards a master/PhD-level student, theoretical and experimental alike, who starts working in the field. Throughout the text, if faced with the choice, we sacrifice mathematical rigor for a more intuitive and concise presentation. There exist a number of recent excellent introductory texts to different subsets of the  subject~\cite{borisreview1,borisreview2,lucas1,levchenko,lucas2,zaanen,geimpolini}. They are either considerably more or less technical than this review. For more technical details and/or mathematical rigor, we refer the readers to Refs.~\cite{borisreview1,borisreview2,lucas1,levchenko,lucas2}. For a non-technical bird's eye view on the subject, we recommend Refs.~\cite{zaanen,geimpolini}.

\subsection{Outline}

In this paper, we distinguish three different scenarios of hydrodynamic behavior in metallic systems: (I) Hydrodynamics in systems close to charge neutrality with a mixture of electrons and holes as elementary excitations, such as graphene. We refer to the hydrodynamics in those systems as electron-hole plasma hydrodynamics, henceforth EHPH; (II) Hydrodynamics in conventional Fermi-liquids, henceforth called FLH; (III) Hydrodynamics in systems with perfect electron-phonon drag, henceforth called Fermi-liquid-phonon hydrodynamics, FLPH. It is important to note that there is in principle also room for electron-hole plasmas that are drag coupled to phonons or even to other collective modes.

In Sec.~\ref{Sec:Hydro}, we discuss generic features of hydrodynamics in metallic systems, the general obstacles, and how they can be overcome. In Sec.~\ref{sec:Boltzmann}, we introduce the Boltzmann equation, our primary framework used throughout this paper. We set it up in a generic manner that allows to describe the formerly mentioned three versions of hydrodynamics in Sec.~\ref{sec:scenarios}: EHPH, FLH, and FLPH. We discuss the respective setups in Sec.~\ref{subsec:EHPH}, Sec.~\ref{subsec:FLH}, and Sec.~\ref{subsec:FLPH} and comment and their most prominent signatures (or at least the ones that are accessible in experiments at the moment). Afterwards, we discuss the experimental status in Sec.~\ref{sec:experiment} and conclude with a discussion and some open question in Sec.~\ref{sec:conclusion}.

\section{Hydrodynamic behavior in electronic lattice systems}\label{Sec:Hydro}


Concepts of fluid motion were introduced to the study of transport properties of fermionic many-body systems already 70 years ago: Fluid behavior was first observed in liquid $^3$He~\cite{volovik}. The first theoretical description goes back to Abrikosov and Khalatnikov in the late 1950s~\cite{abrikosov}. They understood that liquid $^3$He was an example of the then novel Landau Fermi-liquid and that it exhibits hydrodynamic behavior. The relaxation mechanism in that system is the scattering between fermions, which conserves charge, mass, momentum, and energy. It leads to a length scale that governs the relaxational processes: it is called the inelastic mean free path, $l_{ee}$.
The hydrodynamic description is accurate as long as the system is probed over length scales much larger than $l_{ee}$.

In a Fermi-liquid, the inelastic mean free path diverges at low $T$ as $l_{ee} \propto T_F/T^2$ ($T_F$ is the Fermi temperature), which is a direct consequence of fermion-fermion scattering being strongly suppressed due to phase space constraints. Whereas $T_F$ is of the order of a few kelvin for helium at relevant pressures, it is of the order of $1-4\times 10^4$ kelvin for typical metals, which leads to a strong suppression of electron-electron scattering. 
We will later find, that in systems like graphene close to charge neutrality, the inelastic mean free path does not suffer from the same suppression: $l_{ee}\propto 1/T$. Either way, the important message is that having a higher temperature shortens the inelastic mean-free path $l_{ee}$, and therefore takes the system potentially deeper into the hydrodynamic regime.

\begin{figure}[h]
\includegraphics[width=0.75\linewidth]{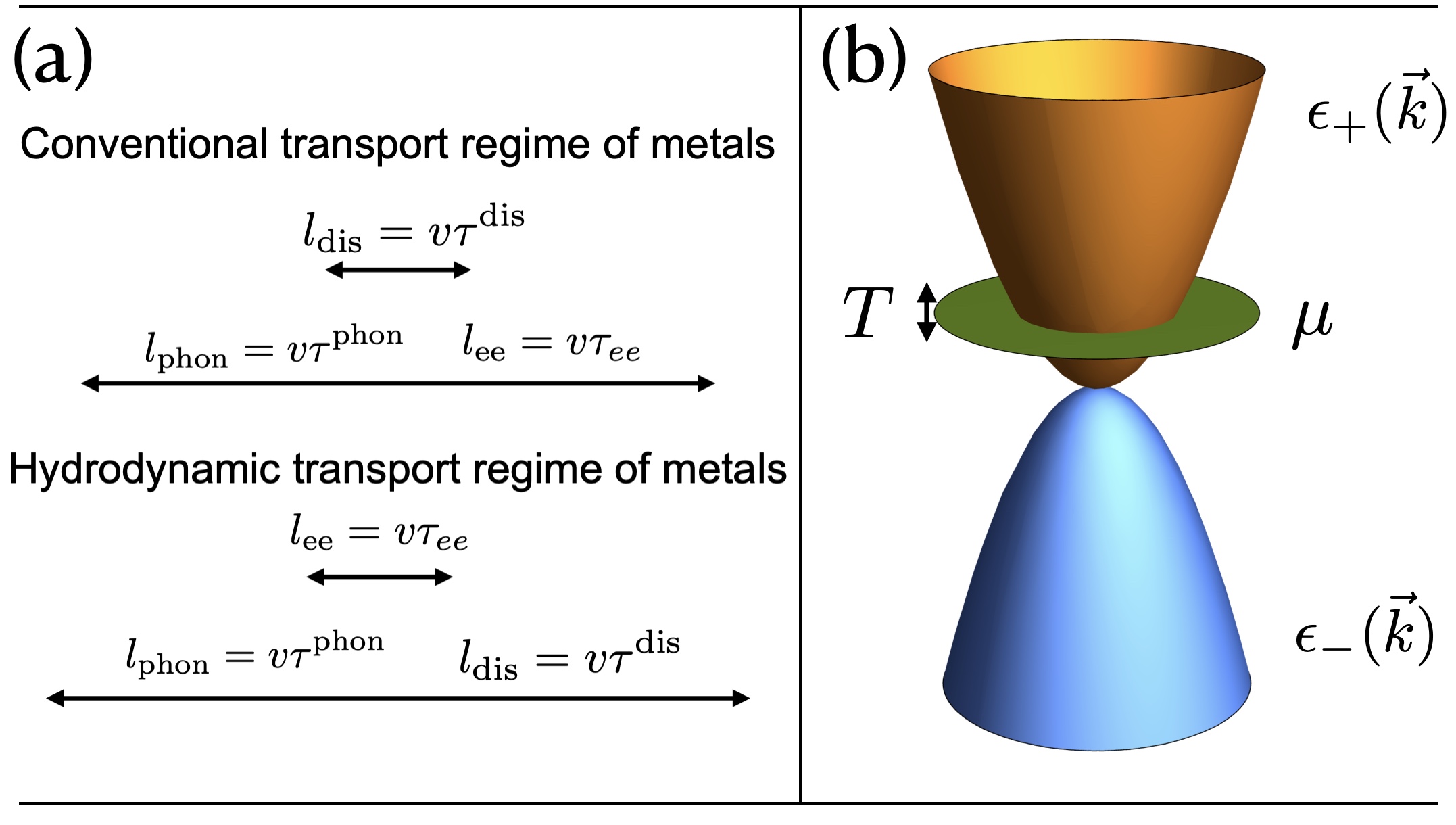}
\caption{(a) In conventional metals relaxation is disorder dominated, whereas in hydrodynamic metals it is interaction dominated. (b) Two-band system consisting of electrons and holes. A chemical potential allows to 'tune' from Fermi-liquid ($|\mu|/T \gg 1$) to an Electron-hole plasma $|\mu|/T \ll 1$. Throughout this paper we assume $\epsilon_+(\vec{k})=-\epsilon_-(\vec{k})$ }
\label{fig2}
\end{figure}

However, typical electronic solid state systems are different from $^3$He in one more crucial aspect: The underlying lattice. It introduces two length/time scales that are absent in $^3$He. One is due to structural disorder within the lattice, called $l_{\rm{dis}}$, and another one is due to scattering from lattice vibrations (phonons), called $l_{\rm{phon}}$. Usually, these length scales, just as previously discussed for $l_{ee}$, are temperature dependent. At low temperatures, electrons mostly scatter from disorder, leading to the textbook residual resistance in metals that is predominantly temperature-independent. At higher temperatures, the main scattering mechanism is due to electron-phonon interactions. In typical three dimensional metals, one finds $l_{\rm{phon}}~\propto 1/T^3$ (or for transport rather $1/T^5$)~\cite{mahan}. Usually, at relevant temperatures, one of these two scattering mechanisms is more effective in restricting electronic motion than electron-electron interaction. This is sketched in Fig.~(\ref{fig2}) (a) where the upper panel shows the conventional metallic regime.

So why do these relaxation mechanisms prevent hydrodynamic behavior? Both disorder and phonon scattering violate conservation laws: Disorder scattering violates momentum conservation, whereas phonon scattering violates momentum and energy conservation.~\footnote{It is important to note that while phonons obstruct hydrodynamics in most systems, it can act as a facilitator in some cases, as discussed explicitly in the FLPH scenario, see Sec.~\ref{subsec:FLPH}.}

In the 1960s, Gurzhi realized that the absence of impurities and phonons is not a strict requirement for hydrodynamic behavior~\cite{gurzhi}. If electron-electron interactions provide the dominant scattering mechanism in a given temperature window, {\it i.e.}, $l_{ee}\ll l_{\rm{phon}},l_{\rm{dis}},W$ ($W$ is the sample width), one can still speak of approximate conservation laws, opening the door for the observation of hydrodynamic behavior, see Fig.~(\ref{fig2}) lower panel.

To summarize, two conditions are favorable to render the lattice sufficiently 'invisible': strong interactions, which can be 'boosted' by increasing temperature, and exceptional sample purity. Concerning phonons there needs to be a high characteristic phonon onset temperature or phonons have to drag-lock with the electrons to form a more complex fluid, as in the FLPH scenario. If all these factors come together, a hydrodynamic window can open at intermediate temperatures.

For decades, lattice systems with those characteristics were not accessible. As a consequence, the field of hydrodynamics received little attention in the study of electronic transport properties in traditional solid state physics. In recent years, however, the situation has improved significantly and very pure materials have become accessible making electronic hydrodynamics an experimental reality.
We will discuss this growing list of systems in Sec.~\ref{sec:experiment}.


\section{Theoretical background}\label{sec:Boltzmann}

 There are two cornerstones of hydrodynamic behavior: conserved quantities and local thermal equilibrium reached through relaxational mechanisms conserving said quantities. Both can be described in the framework of the Boltzmann equation in a very elegant way. This makes the Boltzmann equation our method of choice. In the below section we lay the foundation for all the technical discussions that follow.

\subsection{The setup}

We start with a setup that allows describing all three types of hydrodynamic behavior discussed in this review in one unified framework. 
Our setup consists of electrons ($+$) and holes ($-$) as well as phonons.~\footnote{The phonon could be a collective excitation of the electronic system itself, the formalism looks identical in that case.} Electrons have dispersion $\epsilon_+(\vec{k})$ and holes $\epsilon_-(\vec{k})=-\epsilon_+(\vec{k})$. A chemical potential $\mu$ allows to tune from an electron-hole plasma to a Fermi-liquid at a given temperature $T$: For $|\mu|/T \leq 1$ we have an electron-hole plasma, whereas for $|\mu|/T \gg 1$ we are in a Fermi-liquid limit, see also discussion in Sec.~\ref{subsec:EHPH}. 
Throughout the text, we make the simplifying assumption of isotropy, {\it i.e.}, $\epsilon_{\pm}(\vec{k})=\epsilon_{\pm}(|\vec{k}|)$. The formalism, however, can easily accommodate a more generic dispersion, it only leads to more complicated expressions. A sketch of such a two-band model is shown in Fig.~(\ref{fig2}) (b). The phonons have a dispersion $\omega(\vec{k})$.

\subsection{Boltzmann equation}

The Boltzmann equation is a fundamental equation of statistical physics~\cite{lifshitzpitaevski}. It accommodates the two most important aspects of hydrodynamic behavior: (1) One can easily identify conserved quantities and derive continuity equations; (2) It describes the slow relaxation toward local equilibrium in the presence of conserved quantities. 

At its core, it is a differential equation for the distribution functions of the (quasi-) particles in the system.

We introduce the distribution functions $f_{+} \left(\epsilon_+(\vec{k}),\vec{k}\right)=f \left(\epsilon_+(\vec{k}),\vec{k},\vec{x}\right)$ for the electrons and $f_{-} \left(\epsilon_-(\vec{k}),\vec{k},\vec{x}\right)=f \left(\epsilon_-(\vec{k}),\vec{k},\vec{x}\right)-1$ for the holes.  Subtracting '$1$' from the hole distribution amounts to subtracting the filled lower band. This ensures that we can refrain from using a cutoff when calculating physical quantities from the distribution functions. In equilibrium, the distribution functions for the electrons and holes reduce to the standard Fermi-Dirac distributions $f^0_{\pm}\left(\epsilon_\pm(\vec{k})\right)=\pm \left(\exp \left(\pm \left(\epsilon_{\pm}(\vec{k})-\mu \right)/\left(k_B T\right)\right)+1\right)^{-1}$, where $k_B$ is the Boltzmann constant. Furthermore, we introduce the distribution function $b(\omega(\vec{k}))$ for the phonons. In equilibrium, it is the Bose-Einstein distribution $b^0(\omega(\vec{k}))=\left(\exp \left(\left(\omega(\vec{k})-\mu \right)/\left(k_B T\right)\right)+1\right)^{-1}$. We find three coupled Boltzmann equations (from now on we use $\hbar=k_B=1$ )
\begin{eqnarray}\label{eq:coupledBoltzmann}
\partial_t f_++ \partial_{\vec{k}} \epsilon_+(\vec{k})  \partial_{\vec{r}} f_++  \partial_{\vec{r}} \epsilon_+ (\vec{k})\partial_{\vec{k}} f_+&=&\mathcal{C}^{ee}_++\mathcal{C}^{eh}_+ + \mathcal{C}^{\rm{phon}}_++\mathcal{C}^{\rm{dis}}_+ \;, \nonumber \\
\partial_t f_-+ \partial_{\vec{k}} \epsilon_-(\vec{k})  \partial_{\vec{r}} f_- +\partial_{\vec{r}} \epsilon_- (\vec{k})\partial_{\vec{k}} f_-&=& \mathcal{C}^{hh}_-+\mathcal{C}^{he}_- + \mathcal{C}^{\rm{phon}}_-+\mathcal{C}^{\rm{dis}}_- \;, \nonumber \\ \partial_t b+ \vec{\nabla}_{\vec{k}} \omega(\vec{k})  \partial_{\vec{r}} b+  \partial_{\vec{r}} \omega (\vec{k})\partial_{\vec{k}} b&=&\mathcal{C}^{\rm{int}}_{\rm{phon}} + \mathcal{C}^{+}_{\rm{phon}}+ \mathcal{C}^{-}_{\rm{phon}}\;.
\end{eqnarray}

The left-hand sides are the so-called streaming terms resulting from forces, inhomogeneities, and temporal changes (we neglect Berry-phase terms that are potentially present in two-band systems since we are interested in diagonal response and 'metallic' systems). The right-hand sides describe the collisions of distinct physical origin, all encoded in the collision integrals $\mathcal{C}$. Collisions enable the system to relax towards local thermal equilibrium, a requirement of hydrodynamic behavior. They also couple the three Boltzmann equations and allow the three subsystems, electrons, holes, and phonons, to exchange charge, particles, momentum, and energy. \footnote{The magnetic part of the Lorentz force is included in Eq.~\eqref{eq:coupledBoltzmann} if the momentum is replaced by the canonical momentum according to the minimal coupling prescription~\cite{mahan}.}

\subsubsection{Relaxational processes}
\begin{figure}[h]
\includegraphics[width=0.9\linewidth]{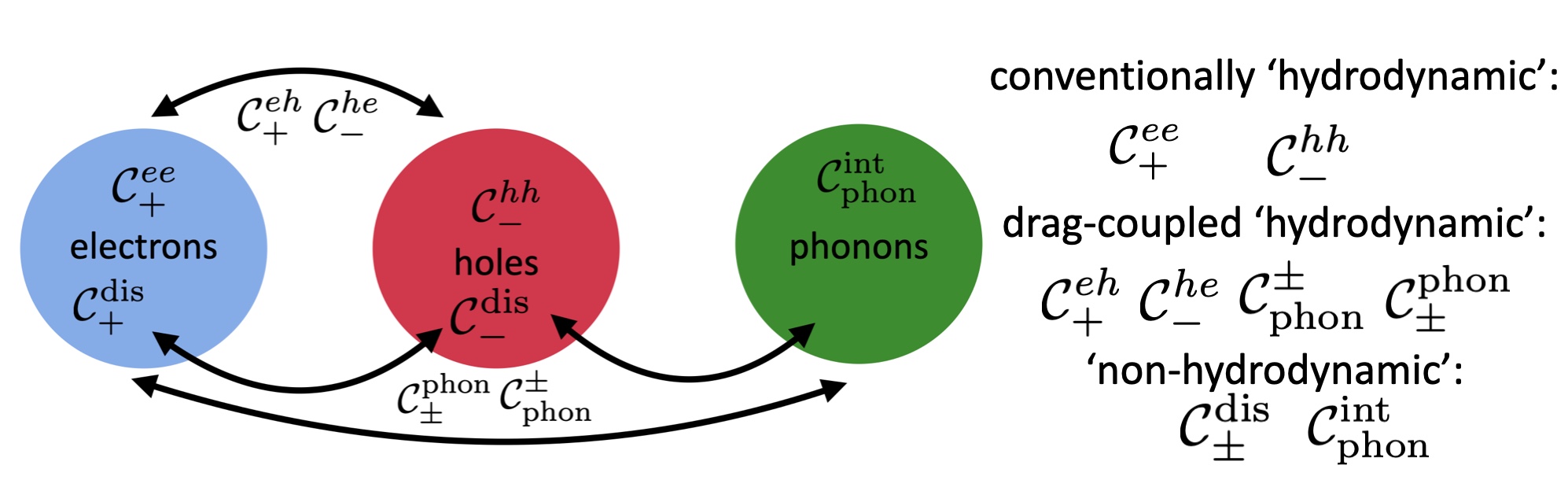}
\caption{Graphical representation of how the different scattering terms act inside and between particle species. The presence or absence of certain terms determines the type of hydrodynamic or conventional transport and flow behavior.}
\label{fig3}
\end{figure}

Eq.~\eqref{eq:coupledBoltzmann} is a coupled set of integrodifferential equations. There are two main difficulties in solving them, all rooted in the collision terms: The collision terms are non-linear and couple all three equations. We split the collision terms into four groups of physically distinct scattering processes:

\begin{enumerate}
\item $\mathcal{C}^{ee}_+$ and $\mathcal{C}^{hh}_-$ describe scattering events in which either only electrons or only holes are involved. The processes conserve the number of electrons and holes, as well as the momentum and energy of both the electron and hole subsystems, respectively. These terms are the terms conventionally associated with hydrodynamic behavior in Fermi-liquids, see Sec.~\ref{subsec:FLH}.

\item $\mathcal{C}^{eh}_+$ and $\mathcal{C}^{he}_-$ describe scattering events between electrons and holes. The processes do not necessarily conserve the 
 number of electrons and holes, individually. Furthermore, they transfer momentum and energy between the electron and hole subsystems. Physically, they correspond to drag terms between the electrons and holes. Overall, the combined system of electrons and holes still conserves total charge, total momentum, and total energy. These terms allow for drag-coupled hydrodynamics in multi-component systems, such as electron-hole plasmas, Sec.~\ref{subsec:EHPH}.

\item $\mathcal{C}^{\rm{phon}}_{\pm}$ and $\mathcal{C}^{\pm}_{\rm{phon}}$ describe the scattering between electrons, holes, and phonons. They allow transferring momentum and energy from electrons and holes to phonons and vice versa. Taken together, they conserve total charge, total momentum, and total energy. Again, these terms allow for drag-coupled hydrodynamics in multi-component systems and are important in Fermi-liquid-phonon setups, Sec.~\ref{subsec:FLPH}.

\item $\mathcal{C}^{\rm{dis}}_{\pm}$ and $\mathcal{C}^{\rm{int}}_{\rm{phon}}$ describe the scattering of electrons and holes from the disorder as well as the internal relaxation of the phonon system. In the case of $\mathcal{C}^{\rm{dis}}_{\pm}$, it conserves the individual number of electrons and holes and consequently total charge as well as energy, but it breaks momentum conservation. The internal phonon term, $\mathcal{C}^{\rm{int}}_{\rm{phon}}$, accounts for a variety of effects: non-linear phonon-scattering, Umklapp scattering, as well as the scattering from disorder in the phonon sector. Potentially, it breaks all conservation laws associated with the phonons. These terms are classified as 'non-hydrodynamic'.
\end{enumerate}
A graphical representation of the role of the aforementioned scattering terms is shown in Fig.~\ref{fig3}. We explicitly distinguish 'hydrodynamic' terms from 'non-hydrodynamic' terms.

One of the critical features of the Boltzmann equation description is that it allows for identifying conserved quantities in a straightforward manner. Formally, one needs to construct the so-called collisional invariants. For a pedagogical discussion of this subject, we recommend consulting Ref.~\cite{lifshitzpitaevski} or similar textbooks about kinetic theory. 

In hydrodynamics, we are always concerned with the conservation of particles/charge, momentum, and energy. The collisional invariants that correspond to these quantities are integrals over the respective quantity and the Boltzmann equation. Concretely, they read$\int \frac{d^dk}{(2\pi)^d}(1,\vec{k},{\rm{Energy}}(\vec{k}))\mathcal{C}=(N,K,E)$ and if a quantity is conserved, they equate to zero. The first integral refers to particle number ($N$), the second to momentum ($K$), and the last one to energy ($E$). For our setup, not all of those quantities have to be zero for the system to be hydrodynamic. Instead, they have to obey sum rules. The quantities are summarized in \textbf {Table \ref{tab1}} for the respective particle type and scattering process.

\begin{table}[h]
\tabcolsep7.5pt
\caption{Collisional invariants}
\label{tab1}
\begin{center}
\begin{tabular}{@{}l|c|c|c@{}}
\hline
&Particle number &Momentum  & Energy  \\
\hline
Electron &$N^{ee}_+,N^{\rm{dis}}_+=0 $& $K^{ee}_+=0$ & $E^{ee}_+,E_+^{\rm{dis}}=0$ \\
 &$N_+^{\rm{phon}},N_+^{eh}\neq 0$& $K_+^{eh},K_+^{\rm{dis}},K_+^{\rm{phon}}\neq0$ & $E_+^{eh},E_+^{\rm{phon}}\neq0$ \\ \hline
Hole  &$N^{hh}_-,N^{\rm{dis}}_-=0$ &$K^{hh}_-=0$ & $E^{hh}_-,E_-^{\rm{dis}}=0$ \\ &$N_-^{\rm{phon}},N_-^{he}\neq 0$& $K_-^{he},K_-^{\rm{dis}},K_-^{\rm{phon}}\neq0$ & $E_-^{he},E_-^{\rm{phon}}\neq0$ \\ \hline Phonon & $N^{\rm{int}}_{\rm{phon}},N^\pm_{\rm{phon}} \neq 0$ & $K^{\rm{int}}_{\rm{phon}},K^\pm_{\rm{phon}} \neq 0$ & $E^{\rm{int}}_{\rm{phon}},E^\pm_{\rm{phon}} \neq 0$ \\
\hline
\end{tabular}
\end{center}
\end{table}
These collisional invariants directly connect to the discussion above and the sketch in Fig.~\ref{fig3}.
We observe that some of the quantities are not generically zero, meaning they correspond to a quantity that is not conserved. To make this very concrete, we consider $K^{eh}_+\neq 0$. This implies that in a collision between electrons and holes, the momentum in the electron sector is not conserved. However, there is a sum rule which reads that $K^{eh}_++K^{he}_-=0$. This implies that the total momentum of the combined system of electrons and holes is conserved, in analogy with the previous discussion. We find a set of six sum rules according to 
\begin{eqnarray}\label{eq:sumrule}
X_+^{eh}+X_-^{he}&=&0 \quad \rm{and} \nonumber \\ X_+^{\rm{phon}}+X_-^{\rm{phon}}+X_{\rm{phon}}^++X_{\rm{phon}}^-&=&0\;,
\end{eqnarray}
where $X$ is $(N,K,E)$. These sum rules can be derived explicitly for rather generic interactions but we content ourselves with interpreting them: (a) The combined system of electrons and holes can exchange particles, momentum, and energy between the subsystems. However, the combined system conserves charge, total momentum, and total energy. (b) The combined system of electrons, holes, and phonons can exchange particles, momentum, and energy between them. Again, charge, phonon number, total momentum, and total energy are conserved in the combined system.
There are only three terms that break these important conservation laws: $\mathcal{C}^{\rm{dis}}_\pm$ breaks momentum conservation of the electrons and holes and $\mathcal{C}_{\rm{phon}}^{\rm{int}}$ which, depending on details, breaks phonon number conservation, momentum, and energy.

\subsubsection{Densities and currents}

Besides the collisional invariants and associated conservation laws, we can also use the Boltzmann equations to derive continuity equations. This can be achieved by integrating the streaming terms of Eq.~\eqref{eq:coupledBoltzmann} over the same quantities.
There, we find densities, currents, forces, and heating terms, see \textbf {Table \ref{tab2}}.
\begin{table}[h]
\tabcolsep7.5pt
\caption{Densities, forces, and Joule heating}
\label{tab2}
\begin{center}
\begin{tabular}{@{}l|c|c|c|c@{}}
\hline
Density &Momentum density & Energy Density & Force  \\
\hline
$n_+= \int f_+$& $\vec{n}^{\vec{k}}_+= \int \vec{k} \;f_+$ & $n^\epsilon_+=\int \epsilon_+ (\vec{k}) f_+$  &$\vec{F}_{+}=\int\dot{\vec{k}}\;f_{+}$ \\ \hline
$n_-= \int f_-$ &$\vec{n}^{\vec{k}}_-= \int \vec{k} \;f_-$ & $n^\epsilon_-=\int \epsilon_- (\vec{k}) f_-$ &$\vec{F}_{-}=\int\dot{\vec{k}}\;f_{-}$ \\ \hline $n_{\rm{phon}}= \int b $ &$\vec{n}^{\vec{k}}_{\rm{phon}}= \int \vec{k} \;b $ & $n^\epsilon_{\rm{phon}}=\int \omega (\vec{k}) b$ &$\vec{F}_{\rm{phon}}=\int\dot{\vec{k}}\;b $\\
\hline
\end{tabular}
\end{center}
\end{table}

In the same way, we find generalized currents shown in \textbf {Table \ref{tab3}}. 
\begin{table}[h]
\tabcolsep7.5pt
\caption{Currents}
\label{tab3}
\begin{center}
\begin{tabular}{@{}l|c|c|c|c@{}}
\hline
Particle Current &Momentum Flux & Energy Current & Heating \\
\hline
$\vec{j}_+=\int \partial_{\vec{k}}\epsilon_+(\vec{k})\; f_+$& $ \Pi^+_{ij}=\int k_i \partial_{k_j}\epsilon_+(\vec{k})f_{+} $ & $\vec{j}^\epsilon_+=\int \partial_{\vec{k}}\epsilon_+(\vec{k})\epsilon_+(\vec{k})\; f_+$ & $h^\epsilon_+= \int \partial_{\vec{k}}\epsilon_+(\vec{k})\cdot \dot{\vec{k}} f_+$ \\ \hline 
$\vec{j}_-=\int \partial_{\vec{k}}\epsilon_-(\vec{k})\; f_-$&$\Pi^-_{ij}=\int k_i \partial_{k_j}\epsilon_-(\vec{k}) f_{-}$ & $\vec{j}^\epsilon_-=\int \partial_{\vec{k}}\epsilon_-(\vec{k})\epsilon_-(\vec{k})\; f_- $ & $h^\epsilon_-= \int \partial_{\vec{k}}\epsilon_-(\vec{k}) \cdot \dot{\vec{k}} f_-$ \\ \hline  $\vec{j}_{\rm{phon}}=\int \partial_{\vec{k}}\omega(\vec{k})\; b$&$\Pi^{\rm{phon}}_{ij}=\int k_i \partial_{k_j}\omega(\vec{k}) b$ & $\vec{j}^\epsilon_{\rm{phon}}=\int \partial_{\vec{k}}\omega(\vec{k})\omega(\vec{k})\; b $ & $h^\epsilon_{\rm{phon}}= \int \partial_{\vec{k}}\omega(\vec{k}) \cdot \dot{\vec{k}} b$ \\
\hline
\end{tabular}
\end{center}
\end{table}

\section{Scenarios of electronic hydrodynamics}\label{sec:scenarios}

\begin{figure}[h]
\includegraphics[width=0.9\linewidth]{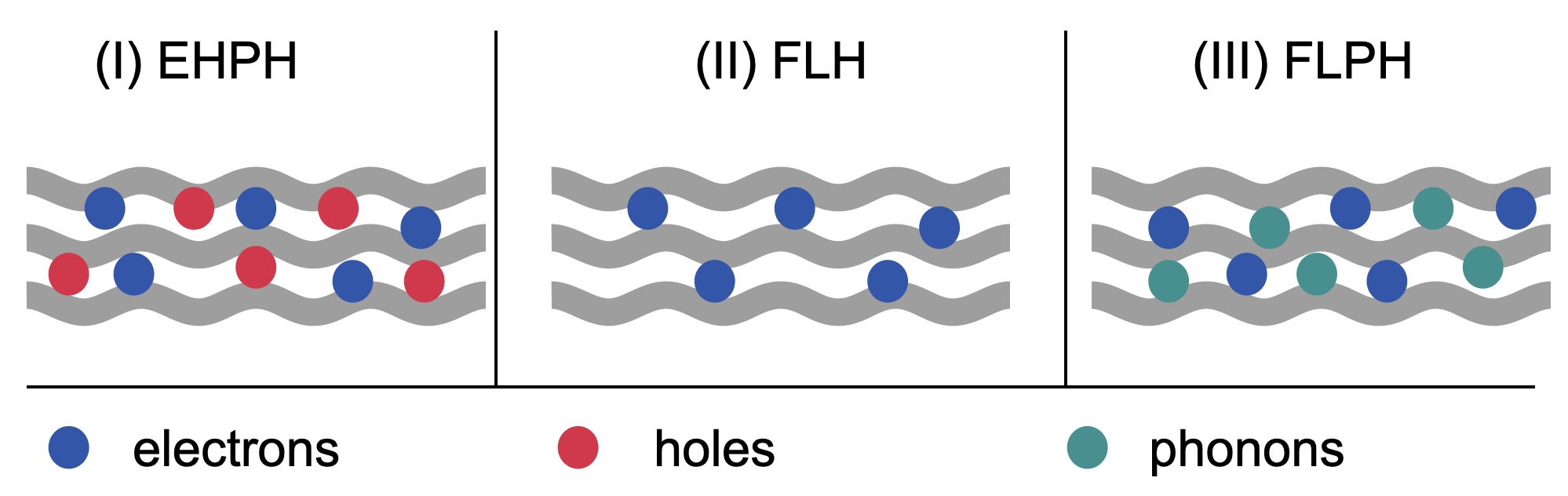}
\caption{Composition of the fluid in the three different scenarios.}
\label{fig4}
\end{figure}

In this review, we consider three different scenarios of electronic hydrodynamics. Our setup can in principle accommodate more complex scenarios, but we focus on three scenarios that are currently most discussed. The three scenarios are: (I) electron-hole plasma hydrodynamics, (II) Fermi-liquid hydrodynamics, and (III) Fermi-liquid-phonon hydrodynamics. The difference between the three scenarios can be made quite pictorial, see Fig.~\ref{fig4}: In (I), the fluid is composed of electrons and holes, in (II) just electrons (or just holes), whereas in (III) it is electrons (or holes) and phonons. The individual constituents in the multicomponent fluid cases (I) and (III) are glued together by the aforementioned drag effects.
In the following we review the three scenarios in some detail using the Boltzmann equation as the workhorse. In each of the scenarios we also discuss one or two key signatures in detail. Those signatures usually also apply to the other scenarios but we assign them in a way which is mostly motivated by the main experiments in the specific group.

\subsection{(I) Electron-hole plasma hydrodynamics (EHPH)}\label{subsec:EHPH}

The prime representatives of the class of electron-hole plasmas are mono- and bilayer graphene in the vicinity of their charge neutrality point. However, all Dirac type systems and potentially even semiconductors at elevated temperatures fall into this category, albeit with modifications. In the following discussion we mainly concentrate on graphene as the most commonly studied system.

\begin{figure}
\includegraphics[width=0.88\textwidth]{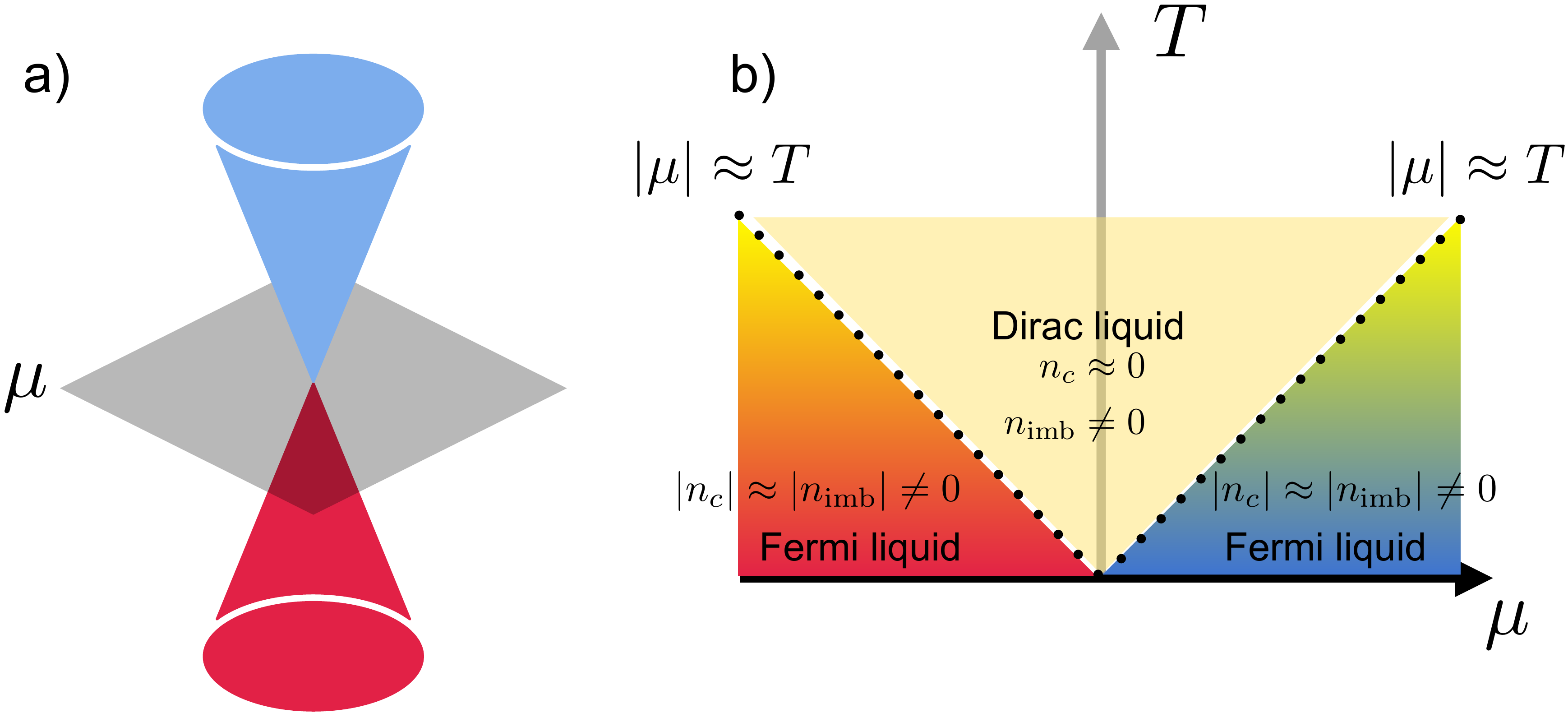}
\caption{a) Schematic of the dispersion relation of graphene near its Dirac point. b) 'Phase diagram' of clean graphene at finite temperature. The region above $\mu=0$ is referred to as Dirac liquid. At $\mu \approx T$ it crosses over to a Fermi-liquid.}\label{fig5}
\end{figure}

Monolayer graphene is a two-dimensional system of carbon atoms on the honeycomb lattice. In its undoped state, it is neither a metal nor an insulator, but a semimetal~\cite{katsnelson,castroneto}. Its density of states is linear in the deviation from the Dirac point. This originates from the low-energy bandstructure, shown in Fig.~\ref{fig5} a). Two bands, of electron and hole type, touch in isolated points in the Brillouin zone. In the vicinity of these points, the system can effectively be described by the massless Dirac equation. Consequently, the spectrum is linear in momentum according to $\epsilon_{\pm}=\pm v_F |\vec{k}|$ where $+$ refers to electrons and $-$ to holes, and $v_F$ is the Fermi velocity, see Fig.~(\ref{fig5}) a). 

Concerning the plasma character of charge-neutral graphene, the key insight came in a paper by Sheehy and Schmalian in 2007~
\cite{sheehy}. The essence is summarized in Fig.~\ref{fig5} b). It shows the 'phase diagram' of graphene as a function of the chemical potential $\mu$ ($x$-axis) and temperature $T$ ($y$-axis). The chemical potential controls the filling of the Dirac cones: The charge density $n_c\propto \mu^2$. Consequently, at $\mu=0$ we have $n_c=0$. However, there are still excitations at finite temperature $T$. A quantity that is sensitive to that is the imbalance density $n_{\rm{imb}}=n_+-n_-$ which behaves according to $n_{\rm{imb}}\propto T^2$. This quantity is a measure for the density of excitations, in that case a thermal cloud of electrons and one of holes, both of equal density, which ensures $n_c=0$. The finite temperature region above $\mu$ has been dubbed the 'Dirac liquid' and it has thermodynamic properties that are very different from Fermi-liquids. The crossover region is defined by the condition $|\mu| \approx T$. For $|\mu| \gg T$, the system behaves like a Fermi-liquid of electron or hole type.\footnote{This discussion is not only valid in graphene, but in any Dirac-type two-band system including bilayer graphene. If the temperature is larger than the respective gap, it even applies to semiconductors.}

The Dirac liquid or Dirac plasma has a number of curious experimental signatures. Some of them become apparent in the bulk thermodynamic quantities, whereas others can be observed in transport probes. Concerning this review, in the case of the electron-hole plasma we mostly concentrate on bulk transport properties. 

\subsubsection{Theoretical description}

The starting point is the Boltzmann equation, Eq.~\eqref{eq:coupledBoltzmann}. To describe the EHPH scenario, we consider the Boltzmann equation of electrons and holes and disregard the contribution due to phonons (a justification of this is mostly of experimental nature and discussed in Sec.~\ref{sec:experiment}), {\it i.e.}, $\mathcal{C}_{\pm}^{\rm{phon}}\approx 0$. The remaining coupled Boltzmann equations read
\begin{eqnarray}
\partial_t f_++ \partial_{\vec{k}} \epsilon_+(\vec{k})  \partial_{\vec{r}} f_++  \partial_{\vec{r}} \epsilon_+ (\vec{k})\partial_{\vec{k}} f_+&=&\mathcal{C}^{ee}_++\mathcal{C}^{eh}_+ +\mathcal{C}^{\rm{dis}}_+ \;,\nonumber \\
\partial_t f_-+ \partial_{\vec{k}} \epsilon_-(\vec{k})  \partial_{\vec{r}} f_- +\partial_{\vec{r}} \epsilon_- (\vec{k})\partial_{\vec{k}} f_-&=& \mathcal{C}^{hh}_-+\mathcal{C}^{he}_- +\mathcal{C}^{\rm{dis}}_- \;.
\end{eqnarray}
The corresponding conservation laws are shown in \textbf {Table \ref{tab4}}.
\begin{table}[h]
\tabcolsep7.5pt
\caption{Conservation laws of electrons and holes}
\label{tab4}
\begin{center}
\begin{tabular}{@{}l|c|c@{}}
\hline
&Electrons & Holes \\
\hline
Particle number &$\partial_t n_+ +\vec{\nabla} \vec{j}_+= N_{+}^{eh} $& $\partial_t n_{-}+\vec{\nabla} \vec{j}_{-}=  N_{+}^{eh}$ \\
Momentum  &$\partial_t \vec{n}^{\vec{k}}_+ +\vec{\nabla}\Pi^+ -\vec{F}_+ = K_+^{eh}+K_+^{\rm{dis}}$ &$\partial_t \vec{n}^{\vec{k}}_{-}+\vec{\nabla}\Pi^{-}-\vec{F}_{-}=K_-^{eh}+K_-^{\rm{dis}}$  \\
Energy &$\partial_t n^{\epsilon}_+ +\vec{\nabla}\cdot \vec{j}^\epsilon_{+}-h^\epsilon_+= E_+^{eh}$ &$\partial_t n^{\epsilon}_{-} +\vec{\nabla} \vec{j}^\epsilon_{-}-h^\epsilon_{-}= E_-^{he}$  \\
\hline
\end{tabular}
\end{center}
\end{table}
It shows that electrons and holes, individually, are not conserved, whereas the total charge density $n_c=n_++n_-$ is. The same statement applies to the energy, where only the total energy $n_c^\epsilon=n_+^\epsilon+n_-^\epsilon$ is conserved, whereas energy can be exchanged between the subsystems. The momentum density is different in that the total momentum, $n_c^{\vec{k}}=n_+^{\vec{k}}+n_-^{\vec{k}}$, is not conserved if disorder is present. In the case of the EHPH, it is useful to introduce imbalance densities according to $n_{\rm{imb}}=n_+-n_-$ and correspondingly momentum and energy. This leads to the continuity equations shown in \textbf {Table \ref{tab5}}

\begin{table}[h]
\tabcolsep7.5pt
\caption{Conservation laws of charge and imbalance}
\label{tab5}
\begin{center}
\begin{tabular}{@{}l|c|c@{}}
\hline
&Charge (I)+(II) &Imbalance (I) \\
\hline
Particle number &$\partial_t n_c+\vec{\nabla} \vec{j}_c=0$& $\partial_t n_{\rm{imb}}+\vec{\nabla} \vec{j}_{\rm{imb}}= N^{ee}_{\rm{imb}}$  \\
Momentum  &$\partial_t \vec{n}^{\vec{k}}_c+\vec{\nabla}\Pi^c-\vec{F}_c=K_c^{\rm{dis}}$ &$\partial_t \vec{n}^{\vec{k}}_{\rm{imb}}+\vec{\nabla}\Pi^{\rm{imb}}-\vec{F}_{\rm{imb}}=K^{eh}_{\rm{imb}}+K^{\rm{dis}}_{\rm{imb}}$  \\
Energy &$\partial_t n^{\epsilon}_c +\vec{\nabla}\cdot \vec{j}^\epsilon_{c}-h^\epsilon_c=0$ &$\partial_t n^{\epsilon}_{\rm{imb}} +\vec{\nabla} \vec{j}^\epsilon_{{\rm{imb}}}-h^\epsilon_{\rm{imb}}= E^{eh}_{\rm{imb}}$  \\
\hline
\end{tabular}
\end{center}
\end{table}

\subsubsection{The thermoelectric response of EHPH}

One of the hallmarks of the hydrodynamic behavior of EHPH can be seen in the bulk thermoelectric response. We will discuss below that it has two key properties: an interaction-dominated bulk electric conductivity (impossible in Fermi-liquids) and an extreme violation of the Wiedemann-Franz law~\cite{mahan}. 

The thermoelectric response of a system is the combined response of the system to an applied electric field $\vec{E}$ and a temperature gradient $\vec{\nabla}{T}$ captured by the thermoelectric response tensor according to~\cite{mahan}
\begin{eqnarray}\label{eq:thermoelectrictensor}
\left( \begin{array}{c} \vec{j}_c \\ \vec{j}^Q \end{array} \right)=\left( \begin{array}{cc} \sigma & \alpha \\ T \alpha & \overline{\kappa} \end{array} \right)\left( \begin{array}{c} \vec{E} \\ -\vec{\nabla} T\end{array} \right)\;.
\end{eqnarray} 
Below we will discuss how to calculate the response coefficients $\sigma$, $\alpha$, and $\overline{\kappa}$.

\paragraph{The relaxation time approximation}\label{sec:relaxationtime}

Solving the Boltzmann equation is very tedious: The collision terms are integral expressions involving the distribution functions themselves. Usually, there is no analytical solution, apart from in thermal equilibrium. Here, we consider near-equilibrium transport phenomena. Those can usually be described within linear-response theory which allows to make progress and eventually leads to solving linear equations. To that end, one linearizes the distribution functions according to $f_{\pm} \approx f^0_{\pm}+\delta f_{\pm}$ and $b \approx b^0+\delta b$ where $\delta f_{\pm}$ and $\delta b$ are deviations from equilibrium. These deviations are linear in the applied perturbations. In the case of thermoelectric transport, Eq.~\eqref{eq:thermoelectrictensor}, those are the field $\vec{E}$ and the temperature gradient $\vec{\nabla} T$. The solution of the linearized coupled Boltzmann equations, while standard, is still a rather technical exercise that usually requires mode-expansion involving potentially complicated numerics~\cite{ziman}. While this is required in some situations, for instance discussed in Sec.~\ref{subsec:FLH}, we can here use the relaxation time approximation~\cite{lifshitzpitaevski}. In this approximation, all the specifics of the collision process are condensed into one quantity: The relaxation time $\tau$.

In general, the relaxation time approximation violates conservation laws. For the purpose of this discussion, we set it up such that it respects all the conservation laws and collisional invariants introduced in  \textbf {Table \ref{tab4}} and \textbf {Table \ref{tab5}} in a straightforward way. Furthermore, we checked that it reproduces the characteristic qualitative features of an actual numerical solution. 

Assuming an applied electric field $\vec{E}$ and a temperature gradient $\vec{\nabla}T$ we find linearized Boltzmann equations in the relaxation time approximation according to

\begin{eqnarray}\label{eq:coupledBoltzmannrelaxtime}
\partial_t \delta f_+ - \vec{\nabla}T\frac{\epsilon_+-\mu}{T}  \vec{\nabla}_{\vec{k}}f^0_+  - e \vec{E} \vec{\nabla}_{\vec{k}}f^0_+&=&-\frac{\delta f_+}{\tau_{+-}}+\frac{\delta f_-}{\tau_{-+}}-\frac{\delta f_+}{\tau^{\rm{dis}}} \nonumber \\ \partial_t \delta f_-  - \vec{\nabla}T\frac{\epsilon_--\mu}{T}  \vec{\nabla}_{\vec{k}}f^0_- - e \vec{E} \vec{\nabla}_{\vec{k}}f^0_+&=&-\frac{\delta f_-}{\tau_{-+}}+\frac{\delta f_+}{\tau_{+-}}-\frac{\delta f_-}{\tau^{\rm{dis}}}\;.
\end{eqnarray}

The relaxation times play distinct physical roles: $1/\tau_{+-}$ and $1/\tau_{-+}$ refer to electron-hole drag, mediated by interactions (we specify this in Sec.~\ref{sec:experiment}), whereas $1/\tau^{\rm{dis}}_{\pm}$ accounts for disorder scattering of electrons and holes, respectively.\footnote{From Eq.~\eqref{eq:coupledBoltzmannrelaxtime} one can explicitly check that the sum rules, Eq.~\eqref{eq:sumrule}, hold.}

Using the expressions introduced in \textbf {Table \ref{tab3}}, we can formulate the charge current $\vec{j}_c$ and energy current $\vec{j}^\epsilon$ (the heat current follows from this according to $\vec{j}^Q=\vec{j}^\epsilon-\mu/e \vec{j}_c$). 
To that end, we combine both Boltzmann equations and assume momentum-independent scattering times. We exploit $\vec{\nabla}_{\vec{k}}\epsilon_+=-\vec{\nabla}_{\vec{k}}\epsilon_-$ and $\vec{\nabla}\epsilon_+=\vec{\nabla}\epsilon_-$ and integrate the Boltzmann equation. In total, we find 
\begin{eqnarray}
\partial_t \vec{j}_c-\vec{\nabla}T \mathcal{T}_c-e\vec{E}\mathcal{E}_c&=& -\frac{\vec{j}_c}{\tau_+} - \frac{\vec{j}_{\rm{imb}}}{\tau_-} -\frac{\vec{j}_c}{\tau^{\rm{dis}}}\;,\nonumber \\ \partial_t \vec{j}_{\rm{imb}}-\vec{\nabla}T\mathcal{T}_{\rm{imb}}-e\vec{E}\mathcal{E}_{\rm{imb}}&=& -\frac{\vec{j}_{\rm{imb}}}{\tau^{\rm{dis}}}\;.
\end{eqnarray}
We can do the same for the energy current, which leads to
\begin{eqnarray}
\partial_t \vec{j}^{\epsilon}-\vec{\nabla}T\mathcal{T}^{\epsilon}-e\vec{E}\mathcal{E}^{\epsilon}&=&-\frac{\vec{j}^\epsilon}{\tau^{\rm{dis}}} \;,\nonumber \\ \partial_t \vec{j}^{\epsilon}_{\rm{imb}}-\vec{\nabla}T\mathcal{T}^{\epsilon}_{\rm{imb}}-e\vec{E}\mathcal{E}^{\epsilon}_{\rm{imb}}&=&  -\frac{\vec{j}^\epsilon}{\tau_-}-\frac{\vec{j}^\epsilon_{\rm{imb}}}{\tau_+}-\frac{\vec{j}^{\epsilon}_{\rm{imb}}}{\tau^{\rm{dis}}}\;.
\end{eqnarray}
\begin{table}[h]
\tabcolsep7.5pt
\caption{Driving term integrals}
\label{tab6}
\begin{center}
\begin{tabular}{@{}l|c|c@{}}
\hline
& Electric Field &Thermal gradient \\
\hline
Electrical & $\begin{array} {c} \mathcal{E}_c=\int_{\vec{k}} {\vec{\nabla}_{\vec{k}}} \epsilon_+ {\vec{\nabla}_{\vec{k}}} \left(f_+^0-f_-^0 \right)  \\  \mathcal{E}_{\rm{imb}}=\int_{\vec{k}} {\vec{\nabla}_{\vec{k}}} \epsilon_+ {\vec{\nabla}_{\vec{k}}} \left(f_+^0+ f_-^0 \right)  \end{array}$ & $\begin{array} {c} \mathcal{T}_c=\int_{\vec{k}} {\vec{\nabla}_{\vec{k}}} \epsilon_+  \left(E_+ {\vec{\nabla}_{\vec{k}}} f_+^0-E_- {\vec{\nabla}_{\vec{k}}} f_-^0 \right)   \\  \mathcal{T}_{\rm{imb}}=\int_{\vec{k}} {\vec{\nabla}_{\vec{k}}} \epsilon_+ \left(E_+ {\vec{\nabla}_{\vec{k}}} f_+^0+E_- {\vec{\nabla}_{\vec{k}}}  f_-^0 \right)   \end{array}$ \\
\hline
Thermal & $\begin{array} {c} \mathcal{E}^\epsilon_c=\int_{\vec{k}} {\vec{\nabla}_{\vec{k}}} \epsilon_+ \epsilon_+ {\vec{\nabla}_{\vec{k}}} \left(f_+^0+f_-^0 \right)  \\  \mathcal{E}^\epsilon_{\rm{imb}}=\int_{\vec{k}} {\vec{\nabla}_{\vec{k}}} \epsilon_+ \epsilon_+ {\vec{\nabla}_{\vec{k}}} \left(f_+^0-f_-^0 \right)  \end{array}$ & $\begin{array} {c} \mathcal{T}^\epsilon=\int_{\vec{k}} {\vec{\nabla}_{\vec{k}}} \epsilon_+ \epsilon_+  \left(E_+ {\vec{\nabla}_{\vec{k}}} f_-^0+E_- {\vec{\nabla}_{\vec{k}}}  f_-^0 \right)  \\  \mathcal{T}^\epsilon_{\rm{imb}} =\int_{\vec{k}} {\vec{\nabla}_{\vec{k}}} \epsilon_+ \epsilon_+  \left(E_+ {\vec{\nabla}_{\vec{k}}}  f_-^0-E_+ {\vec{\nabla}_{\vec{k}}} f_-^0 \right)  \end{array}$ \\
\hline
\end{tabular}
\end{center}
\end{table}

In the above expressions we introduced $1/\tau_+=1/\tau_{+-}+1/\tau_{-+}$, $1/\tau_-=1/\tau_{+-}-1/\tau_{-+}$. The quantities $\mathcal{E}$ and $\mathcal{T}$ can be obtained from straightforward integrals over the Boltzmann equation and are shown in \textbf{Table}~(\ref{tab6}) (note that for notational convenience we have introduced $E_\pm=\epsilon_{\pm}-\mu$). The hydrodynamic limit is reached for $\frac{1}{\tau_+} \gg \frac{1}{\tau^{\rm{dis}}}$ (this is equivalent to $l_{ee}\ll l_{\rm{dis}}$). One can bring these four equations into the more conventional form, Eq.~(\ref{eq:thermoelectrictensor})
\begin{eqnarray}\label{eq:thermoelectricresponse}
\left( \begin{array}{c} \vec{j}_c \\ \vec{j}^Q \end{array} \right)=\left(   \begin{array}{cc} \frac{e\mathcal{E}_c+\frac{\tau_+}{\tau_-} e\mathcal{E}_{\rm{imb}}}{-i\omega+\frac{1}{\tau_+}} +\frac{\tau_+}{\tau_-}\frac{e\mathcal{E}_{\rm{imb}}}{-i\omega+\frac{1}{\tau^{\rm{dis}}}}  &  \frac{\mathcal{T}_c+\frac{\tau_+}{\tau_-} \mathcal{T}_{\rm{imb}}}{-i\omega+\frac{1}{\tau_+}} +\frac{\tau_+}{\tau_-}\frac{\mathcal{T}_{\rm{imb}}}{-i\omega +\frac{1}{\tau_{\rm{dis}}}}  \\  T\left(  \frac{\mathcal{T}_c+\frac{\tau_+}{\tau_-} \mathcal{T}_{\rm{imb}}}{-i\omega+\frac{1}{\tau_+}} +\frac{\tau_+}{\tau_-}\frac{\mathcal{T}_{\rm{imb}}}{-i\omega +\frac{1}{\tau_{\rm{dis}}}}  \right)  &\frac{\mathcal{T}^\epsilon-\frac{\mu \tau_+}{e\tau_-} \mathcal{T}_{\rm{imb}}}{-i\omega+\frac{1}{\tau_{\rm{dis}}} } -\frac{\mu}{e} \frac{\mathcal{T}_c+\frac{\tau_+}{\tau_-} \mathcal{T}_{\rm{imb}}}{-i\omega+\frac{1}{\tau_+}}   \end{array} \right)\left( \begin{array}{c} \vec{E} \\ \vec{\nabla}T \end{array} \right)
\end{eqnarray} 
We are interested in two particular transport coefficients: the electrical conductivity $\sigma$ and the thermal conductivity $\kappa=\overline{\kappa}-T\alpha^2 \sigma$. The latter corresponds to $\vec{j}^Q=-\kappa \vec{\nabla}T$ under the condition of no electric current flow. 

\paragraph{The hydrodynamic electrical conductivity}

\begin{figure}[h]
\includegraphics[width=0.88\textwidth]{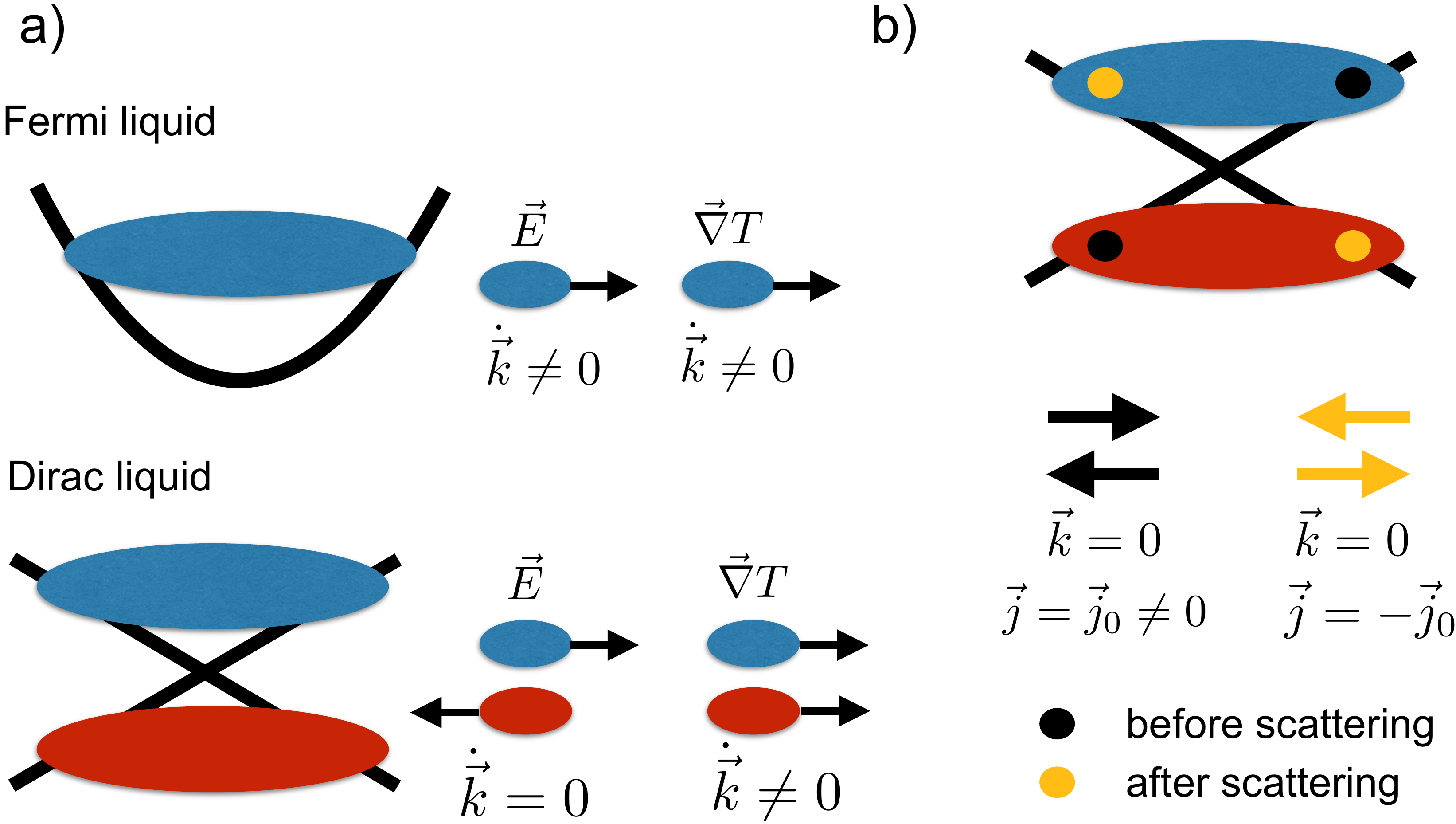}
\caption{a) In a Fermi-liquid, an applied electric field and well as a temperature gradient excite finite momentum. In the Dirac liquid, a temperature gradient excites a finite momentum, whereas an electric field does not. b) In the Dirac liquid, momentum and current decouple. One can relax current without relaxing momentum.}\label{fig6}
\end{figure}

A key signature of electron-hole plasmas is their electrical conductivity which becomes most apparent at charge neutrality. It revolves around a seemingly paradoxical situation. The system has a total charge zero, $n_c=0$. Nevertheless, there is a finite d.c. conductivity. Most surprisingly, this even holds true in the clean limit with no disorder at all. On the level of Eq.~(\ref{eq:thermoelectricresponse}) and \textbf{Table}~(\ref{tab6}) this has the following origin: $\mathcal{E}_{\rm{imb}}=0$ and $1/\tau_-=0$. On the other hand, $\mathcal{E}_c \neq 0$, which implies
\begin{eqnarray}
\sigma_{d.c.}(\mu=0,T)=e^2\mathcal{E}_c \tau_+
\end{eqnarray}
where $1/\tau_+$ is the inverse drag scattering time. As mentioned, this expression is finite even in the absence of disorder, {\it i.e.}, for $1/\tau^{\rm{dis}}=0$. 
The key to understanding this situation is summarized as a sketch in Fig.~\ref{fig6} a) and b).

At the Dirac point, lower panel Fig.~\ref{fig6} a), the charge density $n_c=0$. However, the imbalance density is $n_{\rm{imb}}\neq 0$: At finite temperature, there are two thermal clouds of equal density, one of the electrons and one of the holes. Thus, the total charge is zero. However, an applied electric field pulls electrons and holes in opposite directions. Consequently, the total momentum of the system remains zero, but there is a current induced. Since there is no net momentum induced, there is no disorder required to relax momentum. The electric current, on the other hand, can decay. This is sketched in Fig.~\ref{fig6} b) where an electron-hole pair scatters into another electron-hole pair of opposite 'current'. The momentum, as discussed, is unchanged in this process. Overall, interactions provide a drag mechanism between electrons and holes that effectively 'glues' them together and makes them behave as one fluid. This is sufficient to establish a finite electric current, even without disorder. This is markedly different in a Fermi-liquid, see Fig.~\ref{fig6} a) upper panel. There, an applied field induces momentum and current at the same time (in Sec.~\ref{subsec:FLH} we will see that they are directly proportional). This discussion strictly speaking only applies to the charge-neutrality point.

Tuning away from charge neutrality, $\mu \neq0$, both $\mathcal{E}_c,\mathcal{E}_{imb}\neq 0$ become finite. There is also an effect on the scattering times: $1/\tau_{+-}\neq 1/\tau_{-+}$, which implies that $1/\tau_- \neq 0$. Consequently, the d.c. conductivity reads
\begin{eqnarray}
\sigma_{d.c.}(\mu=0,T)=e^2\left( \mathcal{E}_c+\frac{\tau_+}{\tau_-} \mathcal{E}_{\rm{imb}} \right) \tau_+ + e^2\frac{\tau_+}{\tau_-} \mathcal{E}_{\rm{imb}} \tau^{\rm{dis}}\;.
\end{eqnarray}
It is easy to see that this diverges in the absence of disorder ($1/\tau^{\rm{dis}} \to 0$) as one would expect. The reason is that the finite charge density, $n_c \neq 0$ activates the Drude peak which is associated with momentum transport. Consequently, disorder is required to relax the current, just like in a Fermi-liquid. 

\paragraph{The heat conductivity}

Here, we discuss the response $\vec{j}^Q=-\kappa \vec{\nabla}T$ which involves the coefficient $\kappa$ that is not part of Eq.~\ref{eq:thermoelectrictensor}. It is related to heat conductivity in the absence of current flow. At the charge neutrality point, we can again consider Eq.~(\ref{eq:thermoelectricresponse}) in combination with \textbf{Table}~(\ref{tab6}): Realizing that $\mathcal{T}_c=0$, $\mathcal{T}^\epsilon \neq0$, as well as $1/\tau_-=0$ directly leads to
\begin{eqnarray}\label{eq:thermal}
\kappa=\mathcal{T}^\epsilon \tau^{\rm{dis}}\;.
\end{eqnarray}
Just like in the case of the electric current, there is a finite current induced despite having $n_c=0$. The reason is the same as before, $n_{\rm{imb}}\neq 0$. However, contrary to the case of an electric field, a temperature gradient makes both thermal clouds, electrons and holes, diffuse into the same direction. Consequently, it excites momentum (but no electric current). This implies that a finite response coefficient $\kappa \neq 0$ requires a momentum-relaxing process. This is exactly the interpretation of Eq.~\eqref{eq:thermal} which becomes infinite in the clean limit where $\tau^{\rm{dis}} \to \infty$.
This situation is again depicted in Fig.~(\ref{fig6}).

\paragraph{The Wiedemann-Franz ratio}
An important quantity in the study of metals is the ratio between heat conductivity $\kappa$ and the electric conductivity $T \sigma$. The was already established in 1853~\cite{wiedemann} by Wiedemann and Franz. They observed that for a variety of metals, the ratio $\kappa/(T\sigma)$ tends to a constant value at low temperatures. Later, this was called the Lorenz number~\cite{mahan}. It was experimentally found that it is universal and given by
\begin{eqnarray}
L=\frac{\kappa}{T\sigma}=L_0= \frac{\pi^2}{3} \left( \frac{k_B}{e}\right)^2\;
\end{eqnarray}
which was later explained by Sommerfeldt~\cite{mahan}.
Whether a system tends to this value or not is often taken as empirical evidence of whether the system is a Fermi-liquid or not. The intuitive understanding of the universality of this ratio is that both heat and electric currents are transported by the same type of (quasi-)particle. Additionally, both heat and electric current undergo the same relaxational mechanisms. In the case of a standard metal, this means that both heat and electrical current are limited by the same scattering time, $1/\tau^{\rm{dis}}$, which is due to the disorder.

Considering the above situation, we find 
\begin{eqnarray}
L=\frac{\mathcal{T}^\epsilon}{e^2 \mathcal{E}_c} \frac{\tau^{\rm{dis}}}{\tau_+}\;
\end{eqnarray}
at the charge neutrality point. Not only does this ratio diverge for a clean system, but it is also not a universal quantity: in general, one should expect a strong violation of the Wiedemann-Franz law close to charge neutrality as well as a strong variation across different samples. However, it is a very good measure of the relative strength of elastic and inelastic scattering in the system.
To finish this discussion, it is worthwhile mentioning that the bulk thermoelectric response measures properties of the homogenous flow of the degrees of freedom of the fluid. Therefore, it is not related to the viscosity which is sensitive to the friction of adjacent fluid layers moving at different speeds. Consequently, it is not entirely straightforward to express the viscosity in terms of the scattering times introduced above which are tailored to describe the thermoelectric response.

\paragraph{Navier Stokes}

Most experiments on EHPH so far target bulk thermoelectric transport properties and therefore do not directly probe the viscosity (we discuss one exception in Sec.~\ref{sec:experiment}). We proceed to sketch the derivation of the Navier-Stokes equation in an EHPH system. 

In the case of classical hydrodynamics, the Navier-Stokes equation can be derived from the set of equations introduced in \textbf {Table \ref{tab4}}. The missing ingredient is to assume that there is a slow uniform flow $\vec{u}$ of the fluid which is related to a local equilibrium distribution function
\begin{eqnarray}f_{\pm}=\frac{1}{e^{\frac{\epsilon_{\pm}-\mu-\vec{u}\cdot \vec{k}}{T}}+1}\;.
\end{eqnarray} 
Expanding this to linear order in $\vec{u}$, we find $\vec{j}_c=n_c\vec{u}$, $\vec{j}^\epsilon=(n^\epsilon+\Pi^c)\vec{u}$, $\vec{j}_{\rm{imb}}=n_{\rm{imb}}\vec{u}$, and $\vec{j}^\epsilon_{\rm{imb}}=(n^\epsilon_{\rm{imb}}+\Pi^{\rm{imb}})\vec{u}$ which is true for any type of electronic dispersion. However, the specifics of the dispersion enter in the momentum densities $\vec{n}^{\vec{k}}_c=-\vec{u} \int_{\vec{k}} \vec{k} \cdot \vec{k}\left(\partial_{\epsilon_+}f^0_+ + \partial_{\epsilon_-}f^0_- \right)$ and $\vec{n}^{\vec{k}}_{\rm{imb}}=-\vec{u} \int_{\vec{k}} \vec{k} \cdot \vec{k}\left(\partial_{\epsilon_+}f^0_+ - \partial_{\epsilon_-}f^0_- \right)$ which are not a priori related to a specific thermodynamic quantity. This implies that the Navier-Stokes equations must be derived on a case-by-case basis for different dispersions. The general strategy, however, is that one uses the expressions shown in \textbf{Table}~(\ref{tab4}) and closes the system of equations by relating momentum current to either charge or heat currents, or combinations thereof.  

For graphene close to the charge-neutrality point as well as in the FL regime, this procedure is presented in great details in Refs.~\cite{borisreview1,borisreview2,lucas1}. At the time of writing this review, we are not aware of an explicit derivation in the case of bilayer graphene or other dispersions. 

\subsection{(II) Fermi-liquid hydrodynamics (FLH)}\label{subsec:FLH}
Hydrodynamics for Fermi-liquids was first discussed by Abrikosov and Khalatnikov in the context of liquid helium (Ref.~\cite{abrikosov}) and by Gurzhi~\cite{gurzhi} in the context of electrons in solids.
We focus here on the latter case which has attracted renewed interest in the last few years~\cite{gurzhi0,gurzhi,PhysRevB.21.3279,gurzhi2,PhysRevB.49.5038,dejong,PhysRevLett.52.368,PhysRevLett.71.2465,PhysRevLett.77.1143,Spivak20062071,PhysRevLett.106.256804,PhysRevLett.117.166601,PhysRevLett.113.235901,levitov,PhysRevB.92.165433,PhysRevB.93.125410,PhysRevB.94.125427,2016arXiv161209239G,PhysRevB.95.115425,PhysRevB.95.121301,bandurin,crossno,moll,borisreview1,guo,krishna,scaffidi,PhysRevB.97.045105,PhysRevB.97.121404,PhysRevLett.121.176805,PhysRevB.98.165412,PhysRevB.97.121405,gooth,braem,berdyugin,sulpizio,borisreview2,shavit,ku,levchenko,holder,jenkins,Keser2021,gupta,Krebs2021,Hong2020}.

\subsubsection{Boltzmann equation}

The starting point is again Eq.~\ref{eq:coupledBoltzmann}. In a conventional Fermi-liquid, we have $|\mu|/T \gg 1$ and only one type of charge carrier, either electrons or holes. Without loss of generality, we henceforth concentrate on electrons and consequently drop the $\pm$ index. The equation of interest in that case is
\begin{eqnarray}\label{eq:FLBoltzmann}
\partial_t f+ \vec{\nabla}_{\vec{k}} \epsilon(\vec{k})  \partial_{\vec{r}} f+  \partial_{\vec{r}} \epsilon_+ (\vec{k})\partial_{\vec{k}} f=\mathcal{C}^{ee} + \mathcal{C}^{\rm{phon}}+\mathcal{C}^{\rm{dis}}\;.
\end{eqnarray}
We will now discuss some of the properties of this equation. Contrary to the case of EHPH for which the essence of Coulomb scattering could be simplified down to the electron-hole drag term in the relaxation time approximation (see Eq.~\ref{eq:coupledBoltzmannrelaxtime}), in this case we will need to include the electron-electron scattering term and to keep track of its momentum dependence in order to ensure momentum conservation.

Let us adapt the notation slightly and rewrite the Boltzmann equation as
\bea
\partial_t f + \bv_{\bk} \cdot \nabla_{\br} f - |e| \mathbf{E} \cdot \nabla_{\bk} f  = \mathcal{C}[f] 
\eea
where $\mathbf{E}$ is an external electric field, and $\mathcal{C}=\mathcal{C}^{ee} + \mathcal{C}^{\rm{phon}}+\mathcal{C}^{\rm{dis}}$.
In linear response, using  $f = f_0 + \delta f$, the force term is approximated as $\mathbf{E} \cdot \nabla_{\bk} f \simeq \mathbf{E} \cdot \nabla_{\bk} f_0 = \frac{d f_0}{d \epsilon} \mathbf{E} \cdot v_{\bk} $.
This leads to the linearized Boltzmann equation
\bea
\partial_t \chi + \bv_{\bk} \cdot \nabla_{\br} \chi + \frac{|e|\mathbf{E}}{m v_F} \cdot \bv_{\bk}  = \mathcal{C}[\chi] 
\label{eq:Boltzmann_FL}
\eea
where we have defined $\chi$ by $\delta f = \frac{d f_0}{d \epsilon} m v_F \ \chi$ for later convenience.
Note that the mass $m$ is defined here as $m\equiv k_F / v_F$ and thus behaves at $\sqrt{n}$ for graphene, whereas it is a constant for a parabolic band.
In the limit $T \ll T_F$, one approximates $\frac{d f_0}{d \epsilon} \simeq -\delta(\epsilon - \epsilon_F)$. This means $\chi$ only needs to be defined at the Fermi surface.



\begin{figure}[t!]
\includegraphics[width=0.88\textwidth]{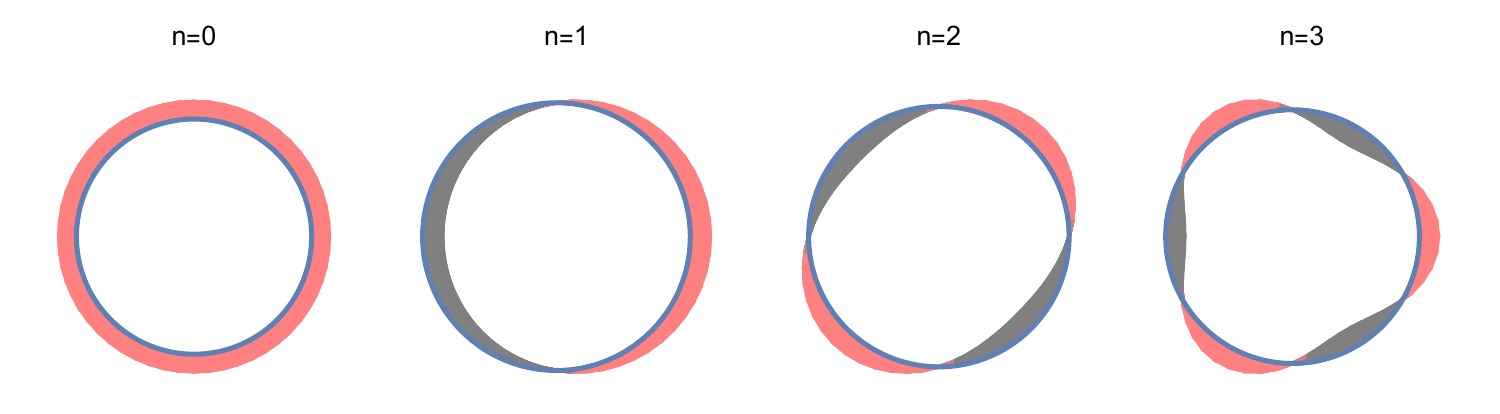}
\caption{Examples of Fourier components ($\chi_0, \chi_{1,x}, \chi_{2,y}, \chi_{3,x}$) contributing to the out-of-equilibrium distribution of electrons at the Fermi surface in a channel geometry. Pink and grey areas denote positive and negative values, respectively.}\label{FSHarmonics}
\end{figure}

For the sake of simplicity, we use the example of a 2D circular Fermi surface in the rest of the discussion.
In this case, one can parametrize the Fermi surface by an angle $\theta$, with $\bk = k_F (\cos(\theta),\sin(\theta))$ and $\bv = v_F (\cos(\theta),\sin(\theta))$.
It is then advantageous to decompose $\chi$ in a Fourier series (see Fig.~\ref{FSHarmonics}),
\bea
\chi(\br,\theta) = \chi_0(\br) + \sum_{n>0}  (\chi_{n,x}(\br) \cos(n \theta) + \chi_{n,y}(\br) \sin(n \theta) ). 
\eea
Three Fourier components are of note: $\chi_0$, which gives the density and is conserved due to charge conservation, and $\chi_{1,x}$ and $\chi_{1,y}$, which give the drift velocity $\vec{u}$ (and thus the current $\vec{j} = ne \vec{u}$) along $x$ and $y$:
\bea
u_{x} = \chi_{1,x}, \  u_{y} = \chi_{1,y}.
\eea
Note that we will use interchangeably current and momentum in this section, since they are proportional in the case of a highly degenerate Fermi-liquid ($T \ll T_F$) with a circular Fermi surface (because $\bv_{\bk} \propto \bk$ for all $\bk$ at the Fermi surface).

Physically, we know that a perturbation $\chi(\theta)$ away from the Fermi-Dirac distribution will tend to decay with time due to scattering.
However, certain Fourier components of $\chi$ might decay faster than others.
This is captured by writing the scattering integral as $\mathcal{C}[\chi] = - \sum_{n>0} \gamma_n \chi_{n} $, where $\gamma_n$ is the decay rate of the $n$-th harmonic.
The set of rates $\gamma_n$ makes it possible to define the scattering integral of any rotationally invariant system, regardless of the microscopic source of scattering.

For concreteness, let us consider a channel of finite width $W$ along $y$ and of infinite length along $x$, where one applies an electric field $\vec{E} = E \hat{x}$.
In the Fourier basis, Eq.~\ref{eq:Boltzmann_FL} takes the form of an infinite set of equations:
\bea
\partial_t \chi_{1} + \frac12 v_F \partial_y(\chi_{2})  &=&  - \gamma_1 \ \chi_{1}  + \frac{e E}{m} \nonumber\\
\partial_t \chi_{2} - \frac12 v_F \partial_y(\chi_{3}-\chi_{1}) &=& - \gamma_2 \ \chi_{2} \nonumber\\
\partial_t \chi_{3} + \frac12 v_F \partial_y(\chi_{4}-\chi_{2}) &=& - \gamma_3 \ \chi_{3}  \nonumber\\
&\vdots & \nonumber\\
\partial_t \chi_{n} - \frac{(-1)^n}2 v_F \partial_y(\chi_{n+1}-\chi_{n-1}) &=& - \gamma_n \ \chi_{n}
\label{eq:Boltzmann_Fourier}
\eea
where $\chi_{n}$ is shorthand notation for $\chi_{n,x}$ (resp. $\chi_{n,y}$) if $n$ is odd (resp. even). This means the contributing Fourier components are $\chi_{1,x}, \chi_{2,y}, \chi_{3,x}, \dots$.
Also, we dropped all the $\partial_x$ terms, since we assume an infinitely long channel.


Naively, one might expect that the only relaxation rate relevant to charge transport is $\gamma_1$, since $\chi_1$ is the mode corresponding to the charge current.
This is indeed the case for a spatially uniform system for which one can neglect all $\partial_y$ terms in Eq.~\ref{eq:Boltzmann_Fourier}. In that case, the first line of Eq.~\ref{eq:Boltzmann_Fourier} is decoupled from the others and gives a closed formula for $\chi_1$ from which the conductivity is found to be $\sigma = n e^2 / m \gamma_1$. We thus recover the Drude formula in that case.


However, in a spatially non-uniform case (e.g. for the finite-momentum conductivity $\sigma(q)$ or for a sample with boundaries), the spatial gradient terms in Eq.~\ref{eq:Boltzmann_Fourier} generate a coupling to higher harmonics. In that case, a knowledge of all the $\gamma_{n>1}$ becomes crucial to understand transport properties~\cite{Nazaryan}.
It is thus the interplay between the spatial non-uniformity and the relaxation of higher harmonics through scattering which leads to interesting effects for Fermi-liquid hydrodynamics. This is in contrast to the electron-hole plasma of the previous section for which bulk properties are already hydrodynamic.

After having motivated the importance of $\gamma_{n>1}$, let us discuss their values.
The simplest approximation is the textbook relaxation time approximation, which assumes a single relaxation rate $\gamma_{n} = \gamma $. 
This assumes that an electron can be scattered anywhere on the Fermi surface with equal probability.
Although this approximation is standard, it actually misses a very important piece of the physics in several important cases.
Notably, electron-electron scattering only contributes to $\gamma_{n>1}$ and not to $\gamma_1$ due to momentum conservation.
This had led to a two-rate model~\cite{abrikosov,Callaway} which separates momentum-relaxing scattering from momentum-conserving scattering:
\begin{eqnarray}
\gamma_1 &=&  \gmr \nonumber\\
\gamma_{n>1} &=& \gmr + \gmc
\end{eqnarray}
where $\gamma_{MR}$ is the momentum-relaxing scattering rate and receives contribution from impurity, phonon, and electron umklapp scattering, where as $\gamma_{MC}$ is the momentum-conserving rate and receives contribution from electron non-umklapp scattering.
When $\gmr \ll \gmc$, one finds a separation of time scales between the slow relaxation of current and the fast relaxation of higher harmonics ($\gamma_1 \ll \gamma_{n>1}$), which justifies a hydrodynamic expansion as explained below.

Remarkably, strong electron-electron scattering is not the only way to realize a substantial separation between $\gamma_1$ and $\gamma_{n>1}$. 
For example, small-angle impurity scattering or certain types of boundary scattering can also lead to a sizable $\gamma_{n>1} / \gamma_1$ ratio, leading to a ``para-hydrodynamic'' regime~\cite{aharon,wolf}.

We should note that even the two-rate model given above is a fairly crude approximation to electron-electron scattering close to a 2D Fermi surface. As shown in Refs.~\cite{ledwith1,ledwith2}, kinetic constraints in 2D lead to an anomalously long lifetime for all odd harmonics, which has important consequences for transport. Further, the special form of the collision integral at the charge neutrality point of graphene also leads to an anomalous kinetic theory\cite{kiselev2}.

\subsubsection{From Boltzmann to Navier-Stokes}

Let us now show how to go from the Boltzmann equation \ref{eq:Boltzmann_Fourier} to the Navier-Stokes equation, using an expansion which relies on the fast relaxation of higher harmonics.
As long as we probe the system at scales much larger than $l_{MC}=v_F \gamma_{MC}^{-1}$, we can do an expansion in a small parameter $\epsilon = l_{MC}/W  $, and use an ansatz of the form $\chi_n \propto \epsilon^{n}$.
To leading order in $\epsilon$, one finds that the first two equations decouple from the rest, giving
\bea
\partial_t \chi_{1} + \frac12 v_F \partial_y(\chi_{2})  &=&  - \gamma_1 \ \chi_{1}  + \frac{eE}{m}  \nonumber\\
 \frac12 v_F \partial_y(\chi_{1}) &=& - \gamma_2 \ \chi_{2} 
\eea
Based on the second equation, we can recognize $\chi_2$ as a component of the viscous stress tensor, which is proportional to the spatial derivative of the flow, as in all Newtonian fluids.
By plugging it into the first equation, we obtain a closed equation for the current $\chi_1$:
\bea
\partial_t \chi_{1} + \eta \partial^2_y(\chi_{1})  &=  - \gamma_\text{1} \ \chi_{1}  + \frac{e}{m} E
\eea
where the viscosity $\eta = \frac14 v_F^2 \tau_2$ is proportional to the relaxation time of the second harmonic, $\tau_2 = 1 / \gamma_2$.
Finally, by using $u_{x} = \chi_{1}$ and generalizing the previous analysis to a two-dimensional flow $\vec{u}$, we find:
\bea
\label{Eq:NSFL}
\partial_t \vec{u} + \eta \nabla^2 \vec{u} = - \gamma_{1} \ \vec{u} + \frac{e}{m} \vec{E}
\eea
which is almost the standard Navier-Stokes equation except for the addition of a momentum-relaxing term $-\gamma_1 \vec{u}$.
The viscous term dominates over the momentum-relaxing one if the channel width $W$ is much smaller than the Gurzhi length: $W \ll \sqrt{l \ l_{MR}}$.
When that is the case, the resistivity is proportional to the viscosity: $\rho \propto \eta / W^2 \propto \tau_2$.~\cite{gurzhi}



By ``integrating out'' the fast microscopic dynamics of the non-conserved higher harmonics $\chi_{n>1}$, we have obtained a closed equation for the slow dynamics of the current $\chi_1$.
The only remaining information about the higher harmonics in this equation is $\tau_2$, which enters the viscosity.

In theory, one could now simply solve Navier-Stokes with the appropriate geometry to study experimental systems, and forget about Boltzmann altogether.
In practice, this is not always justified, since experimental setups do not always have such a large separation of scale between $W$ and $l_{MC}$. 
It is therefore sometimes necessary to solve the full Boltzmann equation.
Further, the question of appropriate boundary conditions for the velocity field also requires going back to a kinetic theory~\cite{kiselev,moessner}.

\subsection{(III) Fermi-liquid-phonon hydrodynamics (FLPH)}\label{subsec:FLPH}

The phenomenon of electron-phonon drag is in principle not restricted to the situation with only electrons and phonons, there could also be a mixture of electrons, holes, and phonons coupled together. In view of the experimental situation, however, we concentrate on electrons and phonons here, the other situation could, however, easily be described within the outlined framework.

This case of hydrodynamic behavior, FLPH, goes against the standard lore of hydrodynamics in electronic solid-state systems: In conventional metals, phonons prevent hydrodynamic behavior. They usually invalidate momentum and energy conservation because they have faster relaxation mechanisms within their subsystem, see discussion below. There is, however, a way out and phonons can act as facilitators that conspire with the electrons and make up one perfectly drag-coupled fluid. The conditions for that are the subject of this section.  

There have been relatively recent review-style articles on the theoretical status of this scenario in Refs.~\cite{levchenko,lucas2}. We keep our discussion more superficial and concentrate on a couple of, in our view, important aspects.

\subsubsection{Theoretical description}
As usual, we start from Eq.~\eqref{eq:coupledBoltzmann}. In a system with electrons and phonons, we have to consider two coupled Boltzmann equations, one for the electrons and one for the phonons (note that we again drop the $\pm$ subscript without loss of generality):
\begin{eqnarray}
\partial_t f+ \vec{\nabla}_{\vec{k}} \epsilon(\vec{k})  \partial_{\vec{r}} f+  \partial_{\vec{r}} \epsilon (\vec{k})\partial_{\vec{k}} f&=&\mathcal{C}^{ee} + \mathcal{C}^{\rm{phon}}_e+\mathcal{C}^{\rm{dis}} \nonumber \\ \partial_t b+ \vec{\nabla}_{\vec{k}} \omega(\vec{k})  \partial_{\vec{r}} b+  \partial_{\vec{r}} \omega (\vec{k})\partial_{\vec{k}} b&=&\mathcal{C}^{\rm{int}}_{\rm{phon}} + \mathcal{C}^{e}_{\rm{phon}}\;.
\end{eqnarray}
The scattering terms in the electron sector fall into three classes: $\mathcal{C}^{ee}$ corresponds to electron-electron scattering that conserves all quantities. $\mathcal{C}_e^{\rm{phon}}$ corresponds to electron-phonon scattering and can transfer momentum and energy between electrons and phonons. Finally, there is $\mathcal{C}^{\rm{dis}}$ which relaxes momentum due to disorder breaking translational symmetry. The scattering terms for phonons fall into two classes: there is scattering between phonons and electrons, called $\mathcal{C}^{e}_{\rm{phon}}$ which is the counterpart to $\mathcal{C}_e^{\rm{phon}}$ and plays the same role. There is also scattering within the phonon system itself, $\mathcal{C}^{\rm{internal}}$. The latter comprises a number of different effects: non-linear terms in the phonon sector, phonon-disorder coupling, and Umklapp scattering. 
In the conventional picture of the electron-phonon problem (see Ref.~\cite{mahan}), $\mathcal{C}^{\rm{int}}_{\rm{phon}}\gg \mathcal{C}^{e}_{\rm{phon}} $. This means the following: phonons relax very quickly on the time scale of electronic processes. Consequently, the phonon system effectively decouples from the electrons and relaxes on its own time scale, usually a lot faster than electronic time scales. This implies that one can treat the phonons as effectively equilibrated in the Boltzmann equation of the electrons. As a consequence, the electronic system has no momentum and no energy conservation since it can dissipate both in the phonon subsystem. Consequently, the electrons will not show hydrodynamic behavior and the transport characteristics are diffusive.

From the point of view of hydrodynamics, there is another very interesting limit: $\mathcal{C}^{\rm{int}}_{\rm{phon}}\ll \mathcal{C}^{e}_{\rm{phon}} $ meaning electron-phonon scattering together with electron-electron interaction determines the equilibration of the combined system. In the clean limit, $\mathcal{C}^{\rm{dis}}=0$, the combined system of electrons and phonons conserves charge, total momentum, and total energy. Hence, the behavior can be expected to be hydrodynamic. Pictorially speaking, the combined system of electrons and phonons is drag-locked into one fluid, see sketch in Fig.~\ref{fig4}. 
 
\subsubsection{Signatures of FLPH}

We again concentrate here on bulk transport properties, especially thermoelectric transport since this gives a rather direct signal. 

\paragraph{Bulk signatures}

The bulk signatures are again best seen in thermoelectric measurements that measure both electric and heat conductivity. There is an intricate competition of effects that comes to life in that situation. First of all, an electric field only couples to the electrons and not the phonons. The electric current is entirely carried by electrons, although phonons are drag-coupled and also move (this is quite similar to the effect of Coulomb drag in double-layer setups). The heat current, on the other hand, is carried by both the electrons and the phonons. Explicitly, we find that the charge current $\vec{j}_c$ and the heat current $\vec{j}_Q$ are given by
\begin{eqnarray}
\vec{j}_c=-e\int_{\vec{k}} \partial_{\vec{k}} \epsilon \delta f \quad {\rm{and}} \quad \vec{j}_Q=\int_{\vec{k}} \partial_{\vec{k}} \epsilon  \left(\epsilon-\mu \right)\delta f +\int_{\vec{k}} \partial_{\vec{k}} \omega\, \omega\, \delta b\;.
\end{eqnarray}
It is important to note that in a Fermi-liquid, the charge current is proportional to the momentum current, see discussion in Sec.~\ref{subsec:FLH}. This implies that an electric field not only excites an electrical current, it immediately excites momentum. The electron-electron interaction, encoded in $\mathcal{C}^{ee}$, consequently is ineffective in relaxing the electrical current. However, electrons can transfer momentum to phonons in the above setup. The phonons themselves cannot dissipate it and all they can do is give it back to the electrons before the cycle repeats. This implies that in the absence of disorder the electrical conductivity is infinite. For a finite electrical conductivity in a Fermi-liquid with perfect phonon drag, one thus needs disorder to relax momentum. To summarize this, the electrical conductivity is expected to the of the form $\sigma=\sigma_0 \tau^{\rm{dis}}_e$ where $\sigma_0$ is a prefactor. 
The thermal conductivity is more complicated than that. The reason is that the thermal current of the electrons is not conserved in inelastic collisions as well as in electron-phonon collisions. The total inverse scattering time of the electronic heat current consists of three terms: $1/\tau^{\rm{eff}}_e=1/\tau^{\rm{dis}}_e+1/\tau^{ee}_e+1/\tau^{\rm{phon}}_e$. There is another direct contribution of the phonons to the heat current. 
Through drag effects, the phonons experience the effects of disorder and electron-electron interaction, meaning the effective scattering time reads $1/\tau^{\rm{eff}}_{\rm{phon}}=1/\tau^{\rm{dis}}_{\rm{phon}}+1/\tau^{ee}_{\rm{phon}}+1/\tau^{\rm{phon}}_{\rm{phon}}$. In total this implies that the Wiedemann-Franz ratio reads
\begin{eqnarray}
L=\frac{\kappa_{0e}\left(\frac{1}{\tau^{\rm{dis}}_e}+\frac{1}{\tau^{ee}_e}+\frac{1}{\tau^{\rm{phon}}_e} \right)^{-1}+\kappa_{0\rm{phon}}\left( \frac{1}{\tau^{\rm{dis}}_{\rm{phon}}}+\frac{1}{\tau^{ee}_{\rm{phon}}}+\frac{1}{\tau^{\rm{phon}}_{\rm{phon}}}  \right)^{-1}}{T \sigma_0 \tau_e^{\rm{dis}}}\;.
\end{eqnarray}
where we defined $\kappa_{0e}$ and $\sigma_0$ such that $L_0=\kappa_{0e}/(T\sigma_0)$ is the Lorenz number.
In the very clean limit, this is given by
\begin{eqnarray}\label{eq:WLFLPH}
L\approx \frac{\kappa_{0e}\left(\frac{1}{\tau^{ee}_e}+\frac{1}{\tau^{\rm{phon}}_e} \right)^{-1}+\kappa_{0\rm{phon}}\left( \frac{1}{\tau^{ee}_{\rm{phon}}}+\frac{1}{\tau^{\rm{phon}}_{\rm{phon}}} \right)^{-1}}{T \sigma_0 \tau_e^{\rm{dis}}}\;.
\end{eqnarray}
While the exact value of $L$ obviously depends on all kinds of details, one must expect that the Wiedemann-Franz law is, possibly strongly, violated.
The two main reasons are: (1) the heat current undergoes relaxational processes differently from the charge current and (2) there is an additional, potentially very big, direct contribution to the heat conductivity coming from the phonons.

It is important to point out that the discussion about the violation of the Wiedemann-Franz law also applies in the case of FLH, Sec.~\ref{subsec:FLH}. In that case the Wiedemann-Franz ratio reads
\begin{eqnarray}
L=\frac{\kappa_{0e}\left(\frac{1}{\tau^{\rm{dis}}_e}+\frac{1}{\tau^{ee}_e} \right)^{-1}}{T \sigma_0 \tau_e^{\rm{dis}}}\;.
\end{eqnarray}
Importantly, this implies that $L<L_0$, as discussed in Refs.~\cite{mahajan,principi}.

\paragraph{Boundary signatures}

In the picture of a strongly coupled electron-phonon fluid, it is obvious that all the quantities involving energy and momentum are heavily influenced by the presence of the phonons. Consequently, the phonons will make a direct contribution to the viscosity, and thus all quantities that are sensitive to viscosity will be different from a Fermi-liquid, even if the Fermi-liquid is hydrodynamic.\footnote{All the mentioned properties of FLPH can also be observed in a scenario of electrons coupled to their own internal collective modes, see Refs.\cite{Kitinan1,Kitinan2,Kitinan3}}.

\section{Experiments}\label{sec:experiment}

In the following discussion we will first discuss the conditions for interactions, disorder, and phonons in some of the relevant systems. Eventually, we discuss the key experiments in the respective groups. 

\subsection{Favorable conditions}

From the discussions in Sec.~\ref{sec:Boltzmann} and the ensuing ones in Secs.~\ref{subsec:EHPH}-\ref{subsec:FLPH} it has become obvious that one key requirement for the experimental observation of solid state hydrodynamics is to have a very clean system (this is important in all three scenarios (I)-(III)). Furthermore, the electrons should be strongly interacting. We will find that this is easier to achieve in scenario (I) than in scenarios (II) and (III). Finally, there is the question about phonons. While in scenarios (I) and (II) their absence is required, in scenario (III) they are explicitly part of the fluid.

\subsubsection{Coulomb interaction}

In all the scenarios, Coulomb interactions are vital to achieve hydrodynamic behavior. The Coulomb interaction is fundamental to all charged fermionic systems. Most importantly, overall it conserves total charge, total momentum, and total energy. Nevertheless, it leads to relaxation of the underlying electronic system. The associated relaxation time, $\tau$, can be very large in Fermi liquids. The reason for that is found in phase space arguments that make relaxation through interactions inefficient, see Ref.~\cite{mahan} or similar textbooks. Without going through the details of the derivation, usually one has $\tau \propto T_F/T^2$ where $T_F$ is the Fermi temperature. This quantity is huge in typical metals, often on the order $1-4\times10^4$ kelvin. However, low-density metals can have a much lower $T_F$, which makes them very attractive. Two of the frontrunners are mono- and bilayer graphene. In those systems, the filling and consequently $T_F$ can be controlled with great precision. It can even be tuned to zero. In that situation temperature itself takes over the role of $T_F$ since it controls the number of thermally excited charge carriers. In that case the scattering time behaves according to $\tau \propto 1/T$. This allows to realize the Planckian limit which is the limit in which relaxation is entirely determined by temperature itself~\cite{subir,hartnollplanck}. 

We will see that also in the other systems that are currently investigated, the electron density is usually quite low (except for a few cases like PdCoO$_2$) which allows to boost the role of Coulomb scattering. Additionally, it is beneficial to consider two-dimensional systems which also increases the relative interaction strength when compared to kinetic energy.

\subsubsection{The role of disorder}

Disorder is the number one enemy of hydrodynamic behavior. Usually, disorder provides the shortest relaxation/scattering time in metals and the only way to increase that is to make the system ever cleaner. There has been tremendous progress in that regard. Especially remarkable is the progress in the field of graphene based devices. Throughout the last six years, it was possible to fabricate encapsulated devices in which mono- or bilayer (or even twisted versions of it) are sandwiched in-between boron-nitrid. This leads to very clean devices in which the effects of disorder scattering can be subdominant to for instance Coulomb scattering, especially if the temperature exceeds $\approx 80$ kelvin.

\subsubsection{The role of phonons}

Lastly, we have to worry about phonons. As explained in detail before, they play a special role in this review. There are two scenarios. One in which phonons equilibrate within their own subsystem and their relaxation is decoupled. In that situation, they are in principle detrimental to hydrodynamic behavior. 
In the case of EHPH and FLH, phonons should be absent. Again, graphene has very favorable properties. Among other things, graphene and bilayer graphene have gained fame for their structural properties. They are very stiff which implies that phonons become important at relatively elevated temperatures around $T=100$ kelvin~\cite{katsnelson}. 
In the scenario of FLPH, phonons fail to relax within their own subsystem and they build a fluid in which the electrons and phonons are locked into one fluid, a scenario which has for instance been discussed in antimony and $\rm{PtSn}_4$.

To summarize this discussion, graphene takes a leading role, mostly due to three properties. (1) $T_F$ can be made very small; (2) Disorder levels can be suppressed and (3) the onset of phonon scattering is above $100$ kelvin. 
Nevertheless, there are also other, possibly three dimensional material systems and we will discuss some of the key experiments below.
{\noindent} Taken together, this leads to an unusual situation. In condensed matter systems, the most spectacular things usually happen upon cooling down the system. Here, the most interesting transport window opens up at elevated temperatures. It has by now been demonstrated by several groups that the hydrodynamic window sits firmly between 10 and 100 kelvin~\cite{ho,tan}. This statement applies to both mono- and bilayer graphene.

\subsection{(I) Electron hole plasma hydrodynamics}

Over the last decade, mono- and bilayer graphene have taken a leading role in the quest for the observation of hydrodynamic electronic flow phenomena of the EHPH type. There have been a number of measurements in which the hydrodynamics of the electron-hole-plasma was probed. 

\subsubsection{The hydrodynamic conductivity}

We discussed in Sec.~\ref{subsec:EHPH} that systems of the EHPH type have a finite electric conductivity even in the perfect system. This is due to the electron-hole drag from Coulomb interactions that enables a current relaxation process. This is a very direct manifestation of hydrodynamics since the hydrodynamic relaxation time is directly related to a measureable bulk quantity. That way it allows a direct measurement of the Planckian dissipation time $\tau \propto 1/T$ (Ref.~\cite{subir}), a quantity that has received a lot of attention in the context of non-Fermi-liquids in recent years (Ref.~\cite{hartnollplanck}). 

\paragraph{Monolayer graphene}

In the case of monolayer graphene, the strength of Coulomb scattering is controlled by a dimensionless coupling constant. In analogy with the fine structure constant in QED, it is denoted $\alpha=e^2/(4\pi \varepsilon v_F)$. In QED, this is a small number, but not in graphene. There are two reasons for that: the Fermi velocity $v_F$ is much smaller than the speed of light, {\it i.e.}, $v_F/c \approx 1/300$, and the dielectric constant $\varepsilon$ depends on the substrate. For practical purposes, this leads to a value of $\alpha=0.2-2$ ($2$ is the extreme case of suspended samples), depending on the details of the setup. The resulting drag scattering rate $1/\tau$ at the Dirac point is found to be of the general form $1/\tau=C\alpha^2T$ where $C=\mathcal{O}(1)$~\cite{kashuba,fritz,foster,mueller,schuett,narozhny,briskot}. The constant $C$ can be calculated from either a solution of the Boltzmann equation or the Kubo formula. Contrary to a standard metal, it is not suppressed by a factor $T/T_F$, where $T_F$ is the Fermi temperature. Gallagher {\it et al.} managed to measure the critical conductivity in monolayer graphene in an optical conductivity measurement and obtained remarkable agreement with the theory results of $\sigma\approx 0.7 e^2/(\alpha^2 h)$~\cite{gallagher}.

\paragraph{Bilayer graphene}

In bilayer graphene in Bernal stacking, the situation is different. This is rooted in the fact that bilayer graphene has a finite density of states at charge neutrality. As a consequence, there is temperature independent Thomas-Fermi screening in the limit of low momenta and the strength of Coulomb interaction does not depend on the fine structure constant. Contrary to the monolayer case, this leads to a temperature independent universal conductivity at the charge neutrality point~\cite{tan,jonathan,vignale,wagner1,wagner2}. Universal in this context means that it is independent of details of the sample.
 To our knowledge, the first direct measurement of the interaction dominated conductivity which matched theory predictions was done in the context of encapsulated bilayer graphene~\cite{nam}. The result was a temperature independent 
interaction limited conductivity of $\sigma \approx 20 e^2/h$ which compared favorably to theory predictions from 2013~\cite{jonathan} (see also~\cite{vignale,wagner1,wagner2,tan}). This was confirmed in a 2019 paper by Tan {\it et al.}. In that paper, it was furthermore shown, that the plasma type physics can even be observed if the bilayer is gapped through the application of an electric field, provided the temperature is large enough. In parallel,  This measurement also allowed to deduce the dielectric constant of the BN-C-BN structure, again in good agreement with theory.

 This has been measured in two independent experiments which are in excellent agreement and also observed this universality across a number of samples~\cite{tan}. As expected from theory, the findings are consistent with a Planckian relaxation time $\tau \propto \hbar/(k_BT )$ which is entirely determined by the constants of nature, $\hbar$ and $k_B$~\cite{subir}.

\subsubsection{Thermoelectric response}

A unique signature of the thermoelectric response of electron-hole plasmas can be found in a maximal violation of the Wiedemann-Franz law, as discussed in Sec.~\ref{subsec:EHPH}. In 2016, Crossno {\it et al.} managed to measure the thermoelectric response of monolayer graphene~\cite{crossno,lucas1} at and in the vicinity of the charge neutrality point. One of the key findings was a strong violation of the Wiedemann-Franz law which was later fit with a two-fluid hydrodynamic theory. In order to find quantitative agreement a careful modelling of the electron-hole puddle disorder structure was required. Nevertheless, this was a strong indication, that the hydrodynamic regime was indeed reached (however, see Ref.~\cite{levchenkoandreev}). Since then, a new series of thermoelectric transport measurement in monolayer and bilayer graphene are under way~\cite{jonah}.

\subsection{(II) Fermi-liquid hydrodynamics (FLH)}
In contrast to the bulk hydrodynamics observed for the electron-hole fluid, one needs to consider finite-size samples to see hydrodynamic effects in transport in the Fermi-liquid regime.
Indeed, since the viscous term in the Navier-Stokes equation (see Eq.~\ref{Eq:NSFL}) is given by $\eta \nabla^2 \vec{v}$, viscosity can only contribute to the resistance when the flow is non-uniform in space, and will reveal itself through size-dependent contributions to transport properties.
The non-local relation between electric field and current means that the sample geometry also has a drastic impact on transport properties.
We will therefore classify experiments according to their geometry.

There is one general caveat though, which is that transport in the ballistic regime (for which all mean free paths are much longer than the sample size) also leads to a non-local relationship between current and electric-field. 
In fact, in both regimes, momentum loss and thus resistance occurs due to boundary scattering.
Hydrodynamics however distinguishes itself by the fact that momentum needs to diffuse through the bulk (which happens microscopically due to frequent momentum-conserving collisions) before reaching the boundary where it can be relaxed.
A challenge for most experiments has therefore been to clearly distinguish between ballistics and hydrodynamics. 


\subsubsection{Channel}
The channel geometry is probably the simplest one and was used in a series of experiments~\cite{molenkamp,dejong,moll,nandi,sulpizio,gooth,ku}.
In this geometry, one can either measure how the conductance $G$ scales with the width $W$ of the channel (with $G \sim W^3$ in the hydrodynamic regime), or use local probes to visualize the Poiseuille flow.
Size-dependent effects in the (negative) magnetoresistance and in the Hall effect also allow in principle a measurement of the shear and Hall viscosities, respectively~\cite{PhysRevLett.117.166601,scaffidi}.
The profile of the Hall electric field across the channel is actually a very useful signature which can discriminate between ballistic and hydrodynamic~\cite{holder}, and which was measured with a local voltage probe (single-electron transistor)~\cite{sulpizio}.

\subsubsection{Widening}
\label{aperture}
In the hydrodynamic regime, injecting current through a narrow aperture into a wide chamber can lead to vorticity laterally from the current.
This vorticity can be measured as a non-local negative resistance\cite{levitov,bandurin,bandurin2}.
A similar geometry was also used to measure the Hall viscosity of graphene~\cite{berdyugin}.
More recently, the emergence of vorticity was also directly imaged with a scanning SQUID, with the appearance of multiple vortices in a circular chamber placed laterally to the flow given as a unique signature of hydrodynamics~\cite{aharon}.
One can even consider more complicated geometries for which a channel forks into several subchannels forming a non-simply connected geometry, like the Tesla valve studied in Ref.~\cite{geurs}.

\subsubsection{Constriction}
A striking hydrodynamic effect is the appearance of superballistic conductance for the flow through constrictions~\cite{krishna,jenkins}.
For a constriction of length $L$ and width $W$, one can distinguish different regimes depending on the scaling of the resistance $R$ with $L$ and $W$.
In the presence of strong momentum relaxation, the Ohmic regime is of course given by $R \propto L/W$.
In the absence of any scattering, one finds the Landauer-Sharvin resistance: $R \propto 1/W$, since the number of conducting channels is proportional to $W$.
Remarkably, adding strong electron-electron scattering to the latter case leads to a hydrodynamic regime for which the resistance is inversely proportional to the constriction length: $R_{hydro} \propto l_{ee}/L W$\cite{stern}.
Superballistic flows were first studied in the case of sharp constrictions ($L \to 0$), for which the resistance goes like $R \sim l_{ee}/ W^2$\cite{krishna,guo}(see also related earlier work in Refs~\cite{nagaev1, nagaev2,melnikov}).
These formulas for the resistance have a simple explanation, since in the hydrodynamic regime, the resistance comes from the viscous term $\eta \nabla^2 \vec{v}$. Whereas the viscosity always leads to a factor of $l_{ee}$, $\nabla^2$ leads to a factor of $1/L W$ or $1/W^2$ for a smooth or sharp constriction, respectively.

\subsubsection{Corbino}
The Corbino (or annular) geometry leads to remarkable manifestations of the hydrodynamic regime of transport.
For currents flowing radially, a Corbino device has effectively no edges (i.e. no boundaries lateral to the flow), which means the usual Poiseuille profile due to non-slip boundary conditions is absent, and the bulk resistance in the hydrodynamic regime is actually zero~\cite{shavit}.
However, the total resistance is actually non-zero due to a voltage drop localized at the contacts which gives a contribution of the type $R \propto \eta / r_{min}^2$, where $r_{min}$ is the smaller radius of the annulus. 

The Corbino geometry makes the physical origin of superballistic flows particularly transparent.
In this geometry, the Landauer-Sharvin resistance is delocalized into the bulk, since the number of channels decreases gradually with the radial coordinates when going from the outer to the inner contact.
In the ballistic regime, only transmitted channels carry current, whereas in the hydrodynamic regime, strong scattering between electrons makes it possible for electrons to hop from reflected to transmitted channels, leading to an increased conductance~\cite{stern,kumarcorbino}.

Additionally, the annular geometry leads to unique effects in thermoelectric~\cite{li} as well as non-linear~\cite{farrell} transport.

\subsubsection{Skin effect}
It is possible to probe the non-local conductivity at varying length scales 
without varying the size of the device by studying AC currents, for which the current density is localized within a frequency-dependent skin depth of the sample edges.
Ohmic, ballistic, and hydrodynamic regimes exist for the skin effect, with differing power laws for the surface resistance dependence on frequency.
A study of these various regimes in PdCoO$_2$ was recently reported in Ref.~\cite{baker}.

\subsubsection{Non-linear transport}

Although there already exists theoretical work on the prospect of reaching higher Reynolds number flows and the instabilities which can result \cite{dyakonov,gabbana,mendl,crabb,petrov,disante,farrell}, only few experiments have studied this non-linear regime so far.
An early example was given in GaAs \cite{dejong}, in which Ohmic heating due a large DC current was used to increase the electronic temperature without increasing the lattice temperature, thereby taking the system deeper in the hydrodynamic regime ($l_{ee} \ll l_{phon}$). 
However, a more recent experiment in graphene showed that electron-phonon coupling can actually become dominant for non-linear transport, leading to a ``phonon Cerenkov'' instability which creates a striking exponential dependence of the resistivity along the current direction~\cite{andersen}.
This electron-phonon instability shows that the physics of non-linear electron hydrodynamics is even richer than previously thought, and certainly deserves further study in the years to come.

\subsubsection{Crossover between FLH and EHPH}
A few experiments have also studied the crossover between the Fermi-liquid and electron-hole plasma regimes, by measuring either channel flow~\cite{ku} or the Wiedemann-Franz ratio ~\cite{crossno}. 


\subsection{(III) Fermi-liquid-phonon hydrodynamics}

As we mentioned before, the FLHP scenario has a very peculiar setup in which electrons and phonons ``lock'' into one fluid. The main proponents of this unusual type of hydrodynamics are currently Sb~\cite{jaoui2,jaoui3}, $\rm{PtSn}_4$~\cite{fu}, and $\rm{WP}_2$~\cite{gooth,jaoui}. This scenario is complicated to prove or rule out in experiments and requires a careful separation of phonon and electron contributions. One theoretical prediction is a strong temperature dependence of the viscosity. Another important feature shows up in thermoelectric measurements together with thermodynamic measurements. According to theory, a hallmark experimental observation that is consistent with an electron-phonon scenario is to see a reduction of the Wiedemann-Franz ratio $L$ below $L_0$, see the discussion in Sec.~\ref{subsec:FLPH}. However, this requires a good knowledge of the phonon contribution to the thermal current.

\section{Conclusion and outlook}\label{sec:conclusion}

Recent years have seen spectacular progress in the study of the hydrodynamic regime in electronic solid state systems. Monolayer and bilayer graphene have taken a leading role but many other systems have emerged. Within this article, we concentrate on three different scenarios of hydrodynamics that are currently most discussed. Electron-hole-plasmas, Fermi-liquids, as well as drag-coupled Fermi-liquid-phonon systems. 
We base the technical parts of the discussion on a phenomenological Boltzmann equation. We devise an easy-to-follow recipe which allows the derivation of the thermoelectric response as well as the Navier-Stokes equation for a relatively generic setup that can accommodate electrons and holes as well as phonons (and in principle other collective modes). The central ingredients in this setup are global conservation laws and drag which transfers momentum and energy between the individual components of the fluid. In the first example, we discussed the electron-hole plasma that is relevant for Dirac-type systems, such as graphene, bilayer graphene, and also Weyl system. One of the main findings is that such systems have finite bulk electric conductivity at charge neutrality, even in the clean limit which is directly related to the relaxation time of hydrodynamic processes, the Planckian time. The second example was that of a strongly coupled electron-phonon system. We also discuss the more conventional Fermi-liquid type hydrodynamics that can best be observed in systems with restricted geometries since they are sensitive to the viscosity. 

In the future, it will be interesting to see to which extent experiments can make smoking gun observations in the respective systems. Ideally, one would like to see real space images of turbulences or other non-linear effects that cannot be explained with more conventional transport theories.

\begin{summary}[SUMMARY POINTS]
\begin{enumerate}
\item A number of recent experiments finds indications for hydrodynamic flow in electronic systems.
\item When comparisons are possible, the agreement between theory and experiment is quite convincing.
\item Monolayer and bilayer graphene are the prime candidates for the observation of hydrodynamic flow phenomena, but many new materials are joining the list.
\item Contrary to common lore, lattice degrees of freedom can help to reach the hydrodynamic electronic limit. 
\end{enumerate}
\end{summary}

\begin{issues}[FUTURE ISSUES]
\begin{enumerate}
\item Can one create and detect turbulence or other nonlinear effects, maybe using novel material systems?
\item What are smoking gun signatures of hydrodynamic behavior?
\item Are there interesting novel effects in the cross-over between Fermi-liquid and electron-hole plasma regime?
\end{enumerate}
\end{issues}

\section*{DISCLOSURE STATEMENT}
The authors are not aware of any affiliations, memberships, funding, or financial holdings that
might be perceived as affecting the objectivity of this review. 

\section*{ACKNOWLEDGMENTS}
We would like to thank numerous collaborators that accompanied and guided us over the years. 
In no particular order, LF acknowledges Subir Sachdev, Markus M\"uller, J\"org Schmalian, Jonathan Lux, Simonas Grubinskas, Kitinan Pongsangangan, Sean Hartnoll, Achim Rosch, Dirk Schuricht, Michael Sch\"utt, Henk Stoof, Matthias Vojta, and Jonah Waissman. This work is part of the D-ITP consortium, a program of the Netherlands Organisation for Scientific Research (NWO) that is funded by the Dutch Ministry of Education, Culture and Science (OCW).
TS acknowledges Andrew Mackenzie, Shahal Ilani and Ady Stern for collaboration on related projects, and the financial support of NSERC, in particular the Discovery Grant [RGPIN-2020-05842], the Accelerator Supplement [RGPAS-2020-00060], and the Discovery Launch Supplement [DGECR-2020-00222].

\end{document}